\documentclass[a4paper,10pt]{article}

\pdfoutput=1

\usepackage{multirow}
\usepackage{amsmath}
\usepackage{tgtermes}
\usepackage{xcolor}
\usepackage{amsfonts}
\usepackage{amssymb}
\usepackage{amstext}
\usepackage{import,breqn}
\usepackage{graphicx}
\usepackage{caption,subcaption}
\usepackage{float}
\usepackage{bm}
\usepackage{makecell}
\usepackage{empheq}
\usepackage{hyperref}
\hypersetup{
	colorlinks=true,
	linkcolor=blue,
	urlcolor=blue,
	citecolor=red
}
\usepackage{epsfig}%
\usepackage{pdfrender,xcolor}
\usepackage{cancel}

\usepackage{authblk}
\usepackage{orcidlink}
\usepackage{cite}

\allowdisplaybreaks[1]

\DeclareMathOperator{\csch}{csch}

\title{\bf A model-independent compact dynamical system formulation for exploring bounce and cyclic cosmological evolutions in $f(R)$ gravity}
\author[1]{Saikat Chakraborty\orcidlink{0000-0002-5472-304X}
	\thanks{\href{mailto:saikat.ch@nu.ac.th}{saikat.ch@nu.ac.th}}}
\author[2]{Charlotte Louw\orcidlink{0009-0009-7835-9631}
	\thanks{\href{mailto:lwxcha023@myuct.ac.za}{lwxcha023@myuct.ac.za}}}
\author[3]{A. S. Agrawal\orcidlink{0000-0003-4976-8769}
	\thanks{\href{mailto:asagrawal.sbas@jspmuni.ac.in}{asagrawal.sbas@jspmuni.ac.in}\ (Corresponding Author)}}
\author[4]{Peter K.S. Dunsby\orcidlink{0000-0002-6271-9585}
	\thanks{\href{mailto:peter.dunsby@uct.ac.za}{peter.dunsby@uct.ac.za}}}
 \affil[1]{The Institute for Fundamental Study, Naresuan University, Phitsanulok 65000, Thailand}
\affil[2,4]{Department of Mathematics and Applied Mathematics, Cosmology and Gravity Group,  University of Cape Town, Rondebosch, 7701, Cape Town, South Africa}
\affil[3]{Department of Mathematics, Jayawant Shikshan Prasarak Mandal University, Pune, India}
\affil[1,4]{Centre for Space Research, North-West University, Potchefstroom 2520, South Africa}
\affil[4]{South African Astronomical Observatory, Observatory 7925, Cape Town, South Africa}

\date{}
	
\begin{document}
	
\maketitle	

\begin{abstract}
Using the dynamical systems approach together with the cosmographic parameters, we present a \emph{model-independent} dynamical system formulation for cosmology in $f(R)$ gravity. The formulation is model-independent in the sense that one needs to specify not a particular functional form of $f(R)$ a-priori, but rather a particular cosmological evolution, which fixes the cosmography. In a sense, our approach is the way around the reconstruction method. This is shown using both non-compact and compact dynamical variables. The focus in this paper is on the compact analysis since we demonstrate the applicability of this formulation using examples of bouncing and cyclic cosmology. In particular, our analysis reveals, in a model-independent manner, the problem of achieving such cosmologies when the universe is globally spatially flat and devoid of matter.
\end{abstract}
	
	

\section{Introduction} 
The standard cosmological model is built upon two key assumptions. Firstly, we assume that General Relativity correctly describes the interaction between the geometry and matter of the universe. Secondly, we assume that on large scales, the universe is homogeneous and isotropic, which is known as the Cosmological principle. While this may seem counter-intuitive, based on our living experience of clumped matter, i.e., inhomogeneity on small scales, we choose to work on a homogeneous and isotropic background and introduce perturbations to account for the large-scale structure that we observe. In order to describe, geometrically, this kind of background, we make use of the Friedmann-Lema\^itre-Robertson-Walker (FLRW) metric. The primary evidence for the Cosmological Principle is the observed uniformity of the temperature of the Cosmic-Microwave Background (CMB)\cite{Ostriker:1995rn}. Thus if we adopt the homogeneity and isotropy as an appropriate description of our universe, we then need to develop our understanding of the gravitational interaction in order to describe both the background expansion history and the evolution of large-scale structure.

General Relativity (GR) has been extremely successful at correctly describing the gravitational interaction on scales ranging from the solar system to extragalactic scales. This is supported by a growing database of observations. However, if one accepts the geometry presented by the FLRW metric, then observational evidence seems to suggest that our universe is undergoing accelerated expansion in the present epoch, thus contradicting the attractive nature of gravity in the standard GR model. There seem to be two lines of thought as to why this is the case. The first proposition involves the existence of ``Dark Energy", a mysterious matter component which currently dominates the energy density of the universe. This component appears to affect gravity in such a way that it becomes repulsive in its presence. While this possibility keeps the GR description of gravity, it leads to the problem of finding a physical description of Dark Energy. A popular candidate is that of the cosmological constant $\Lambda$, which, when added to the Einstein-Hilbert action, presents itself with a constant energy density and equation of state parameter $\omega=-1$. The famous $\Lambda$CDM model of the universe is made up of $\Lambda$ as well as another mysterious component, Cold Dark Matter (CDM)\cite{Ostriker:1995rn}. While it has been shown that this model agrees mostly with observations, it suffers from some problems, particularly the fine-tuning of the energy density of $\Lambda$. The quantum vacuum energy is the only known possibility for describing $\Lambda$. However, its value calculated in quantum field theory differs from the observed value of $\Lambda$ by between 50 and 120 orders of magnitude \cite{Bengochea_2020}.

Thus, we consider the second line of thought attempting to explain our universe's accelerated expansion. While the first approach involved altering the right-hand side of Einstein's equation $G_{\mu\nu}=\kappa T_{\mu\nu}$, one may also consider altering the left side i.e., the acceleration of the universe could be a sign of a break down of the description of the gravitational action according to GR \cite{Capozziello_2008} on cosmological scales. Modified theories of gravity ask the question of whether our understanding of how gravity behaves at large scales is correct. An active area of studies of Modified Gravity involve modifying the Einstein-Hilbert Lagrangian by introducing, in general, two propagating scalar gravitational degrees of freedom in the following form:
\begin{equation*}
R \to f(R,\phi).
\end{equation*}
These degrees of freedom present themselves as an explicit scalar field $\phi$ as well as a hidden propagating degree of freedom given by $F(R,\phi)=\frac{\partial f}{\partial R}$ \cite{Chakraborty2021}. The matter part of the Lagrangian remains as it is. \footnote{From a particle physicist's point of view, adding a scalar field, even in the gravity sector, does change the standard matter content as known in the standard model of particle physics.} 

Of particular interest (both in general relativity and its various extensions) is whether the Universe originated from a singularity at the beginning or began expanding after a non-singular bounce-like phase. Since the inflationary scenario can generate a power spectrum that is almost scale-invariant and in line with the most recent observations, it is an appealing early Universe scenario. Despite inflation's successes, it still depicts a universe that originates from a singularity. In that regard bouncing cosmology idea is an attractive alternative to the inflationary paradigm. In the case of the bouncing scenario, the universe begins in a contracting period, bounces and when the scale factor acquires minimum size, and then begins to expand again. As a result, the bounce happens when $H=0$ and $\dot{H}>0$. In some bouncing cosmology scenarios, in addition to producing an observationally compatible power spectrum, having an expansion history free of an initial singularity is another benefit of such a model.

The modified Friedmann and Raychaudhuri equations, in conjunction with the matter equation of state, make up the closed set of field equations for $f(R)$ gravity. In general, they make up a complicated collection of coupled non-linear partial differential equations. To retain the cosmological principle, we primarily use homogeneous metrics in cosmology, which causes the cosmological field equations to collapse into a collection of coupled nonlinear ordinary differential equations. In order to extract quantitative as well as qualitative information about the evolution of such nonlinear systems, the dynamical system approach can be a useful strategy for studying these equations.  The cosmological field equations can be written as a closed set of coupled first-order nonlinear differential equations, which constitutes the dynamical system, by appropriately defining a set of dimensionless normalised dynamical variables in the dynamical system formulation of cosmology. Dynamical systems can be divided into two categories: continuous dynamical systems, whose evolution is described by a collection of ordinary differential equations and time-discrete dynamical systems, whose evolution is described by a map or difference equations. 

In this paper, we generalise a method first introduced in \cite{Chakraborty2021} in which the reconstruction methods are completely avoided by making use of a dynamical systems approach that does not require the knowledge of the functional form of the $f(R)$ model. The dynamical systems approach has the advantage of providing us with a qualitative understanding of cosmological models and have seen much success in cosmology (see \cite{1999} for an application to the $\Lambda$ CDM model). Almost all past autonomous (not explicitly time-dependent) dynamical system approaches to $f(R)$ models in literature have required the functional form of $f(R)$ to be specified in order to close the system. One can avoid this by introducing a set of dimensionless cosmographic parameters which relate to the kinematical description of the universe \cite{Dunsby:2015ers}. These parameters are obtained by taking a Taylor expansion of the scale factor around the present redshift. As was shown in \cite{Chakraborty2021}, if a kinematical description of the universe is specified, then these parameters define a new constraint which allows the dynamical system to be closed. Therefore, one can solve the system, find the fixed points and study the phase space dynamics without knowledge of the functional form of the $f(R)$ gravity and in fact, use this method to explicitly determine the form of $f(R)$ from the cosmological dynamics.

This cosmography approach is applied to several expansion histories that admit cyclic and bouncing behaviour. By fixing the expansion history, one can easily determine the relationship between the higher-order cosmographic parameters, allowing one to close the system of dynamical equations in a way which is independent of the form of $f(R)$. Our approach is especially well suited to investigate spatially flat and purely curvature driven nonsingular cosmologies, as in this case the phase space becomes 2-dimensional and we can clearly show in the phase space the region where the cosmology is free from any theoretical pathology.

This paper begins by deriving the field equations for a general $f(R)$ gravity under the usual isotropic and homogeneous FLRW background with a perfect fluid. In Section \ref{sec:mod_ind_dyn_sys}, we present the model-independent dynamical systems formulation and compactify the phasespace in Sections \ref{sec:comp_dyn_sys} and \ref{sec:mod_ind_comp_dyn_sys} In Section \ref{sec:applications}, we apply our formalism to three examples where the expansion history is fixed by choosing the explicit form of the scale factor. This allows one to determine the relation between the deceleration and jerk parameters and thus close the system. Finally, in Section \ref{sec:conclusion}, we present our conclusions and comment on possible future work. 
\section{Field equations and the dynamical system for \texorpdfstring{$f(R)$}{} gravity}\label{sec:dyn_sys}
$f(R)$ gravity can be derived from the following action \cite{Sotiriou:2008rp,DeFelice:2010aj}:
\begin{equation}\label{action}
    S = \frac{1}{2\kappa} \int \sqrt{-g} d^{4}x [f(R) + 2\mathcal{L}_{m}]\,,
\end{equation}
 where $\kappa = 8\pi G$, $G$ being Newton's gravitational constant, $g$ is the metric determinant, $f(R)$ is a function of the Ricci scalar $R$, $\mathcal{L}_m$ is the matter action. Varying the action with respect to the metric yields the field equations 
 \begin{equation}\label{field_eq}
G_{\mu\nu} \equiv R_{\mu\nu}-\frac{1}{2}g_{\mu\nu}R = \frac{\kappa T_{\mu\nu}}
{F(R)} + g_{\mu\nu}\frac{\left[f(R)-RF(R)\right]}{2F(R)} + \frac{\nabla_{\mu}\nabla_{\nu}F(R)-g_{\mu\nu}\square F(R)}
{F(R)}\,,
\end{equation}
where the prime denotes a derivative with respect to $R$, $T_{\mu\nu}$ is the energy momentum tensor and $F(R)=f'(R)$. In writing the field equation in the above manner we have, naturally, assumed $F(R)\neq0$, otherwise, there would be no gravity. $f(R)$ theories of gravity propagate an extra scalar degree of freedom in the gravity sector, sometimes dubbed as a `curvaton', as made clear from the trace field equation
 \begin{equation}
    RF(R) - 2f(R) + 3\square F(R) = \kappa T \,,   
\end{equation}
$f(R)=R$ is the trivial case for which the action reduces to GR, in which case the above equation becomes algebraic. Thus, there is no extra propagating scalar degree of freedom. The absence of ghost and tachyonic instability requires $f'(R)=F(R)>0$ and $f''(R)=F'(R)\geq0$ respectively \cite{Sotiriou:2008rp,DeFelice:2010aj}.
 
 For a Friedmann-Lema\^itre-Robertson Walker (FLRW) metric
 \begin{equation}\label{metric}
     ds^{2}=-dt^{2}+a^{2}(t)\left[\frac{dr^{2}}{\left(1-kr^{2}\right)}+r^{2}d\theta^{2}+r^{2}sin^{2}\theta d\phi^{2}\right]\,,
 \end{equation}
 and in the presence of a perfect fluid $T^{\mu}_{\nu} = (-\rho, P, P, P)$, the modified Friedmann and Raychaudhuri equations can be neatly expressed as
\begin{subequations}
    \begin{eqnarray}\label{field_eqs}
        3F\left(H^{2} + \frac{k}{a^{2}}\right) = \kappa \rho_{\rm eff} = \kappa(\rho + \rho_{R})\,,\label{eq:fried}\\
        -F\left(2\dot{H} + 3H^{2} + \frac{k}{a^{2}}\right) = \kappa P_{\rm eff} = \kappa(P + P_{R})\,,\label{eq:Raychoudhuri}
    \end{eqnarray}     
\end{subequations}
where $\rho,\,P$ are the fluid energy density and pressure, and the curvaton energy density and pressure are
\begin{subequations}
    \begin{eqnarray}\label{scalaron_eqs}
        \kappa \rho_{R}&=&\frac{1}{2}(RF-f)-3H\dot{F}\label{eq:sed}\\
        \kappa P_{R}&=&\ddot{F}+2H\dot{F}-\frac{1}{2}(RF-f)\label{eq:sp}
    \end{eqnarray}
\end{subequations}
 Over dot represents the differentiation with respect to cosmic time throughout the paper. The perfect fluid follows the standard continuity equation 
\begin{equation}
    \dot{\rho} = - 3H(1+\omega)\rho \,.   
\end{equation} 
The effective equation of state parameter and the curvaton equation of state parameter are defined, respectively, as
\begin{subequations}
\begin{eqnarray}
&& \omega_{\rm eff}=\frac{P_{\rm eff}}{\rho_{\rm eff}}=-\frac{2\dot{H}+3H^{2}+\frac{k}{a^{2}}}{3H^{2}+\frac{k}{a^{2}}}\,,\label{eff_eos}\\
&& \omega_{R}=\frac{P_{R}}{\rho_{R}}=\frac{\ddot{F}+2H\dot{F}-\frac{1}{2}(RF-f)}{\frac{1}{2}(RF-f)-3H\dot{F}}\,.\label{scalaron_eos}
\end{eqnarray}    
\end{subequations}

The usual dynamical system formulation for cosmology in $f(R)$ gravity in terms of expansion-normalized (or Hubble-normalized) dynamical variables was introduced in \cite{Carloni_2005,Carloni:2007br,Abdelwahab:2007jp} \footnote{For a slightly different formulation, also in terms of Hubble-normalized dynamical variables, see \cite{Amendola:2006we}}. The formulation starts with the introduction of dimensionless dynamical variables
\begin{eqnarray}\label{nc_dyn_variable}
    x = \frac{\dot{F}}{HF}, ~~~
    y = \frac{R}{6H^{2}}, ~~~
    z = \frac{f}{6FH^{2}}, ~~~
    \Omega = \frac{\kappa \rho}{3FH^{2}}, ~~~
    K = \frac{k}{a^{2}H^{2}}\,.    
\end{eqnarray}
They are obtained by dividing the modified Friedmann equation \eqref{eq:fried} by $3FH^{2}$, and satisfy the constraint
\begin{equation}\label{nc_constraint}
    - x + y - z + \Omega - K = 1\,.
\end{equation}
Note that the above definitions of the dimensionless dynamical variables are viable only for $F\neq0$ and $H\neq0$. Introducing on the phase space a dimensionless time variable $\tau =\epsilon\ln a$ ($\epsilon=H/|H|=\pm1$ for expanding/contracting universe) and the auxiliary quantity 
\begin{equation}\label{Gamma}
    \Gamma(R) \equiv \frac{d \ln R}{d \ln F} \equiv \frac{F}{RF'}\,,
\end{equation}
the dynamical system can be written as
\begin{subequations}\label{nc_dy_sys}
    \begin{eqnarray}
        \frac{dx}{\epsilon d\tau}&=&-4z-2x^{2}-(z+2)x+2y+\Omega(x+1-3\omega)\,,\label{eq:nc_x} \\  
        \frac{dy}{\epsilon d\tau}&=&y[2\Omega-2(z-1)+x(\Gamma-2)]\,,\label{eq:nc_y} \\
        \frac{dz}{\epsilon d\tau}&=&z(-2z+2\Omega-3x+2)+xy\Gamma \,,\label{eq:nc_z} \\
        \frac{d\Omega}{\epsilon d\tau}&=&-\Omega(3x+2(z-\Omega)+(3\omega+1))\,,\label{eq:nc_Omega}
    \end{eqnarray}    
\end{subequations}
where $K$ has been eliminated using the constraint \eqref{nc_constraint}. For a given functional form of $f(R)$, one can write the relation 
\begin{equation}
    \frac{y}{z}=\frac{RF}{f}\,.   
\end{equation}
To make the dynamical system autonomous, it is necessary to invert the above relation to find $R=R(y/z)$, provided, of course, that it is invertible. When this is possible, one can close the dynamical system by writing $\Gamma(R) = \Gamma(R(y/z))$. In practice, this is possible only for very specific forms of $f(R)$.

Under the assumption $H\neq0$ that underlies the choice of the dimensionless dynamical variables in Eq.\eqref{nc_dyn_variable}, cosmological solutions with vanishing Ricci scalar form an invariant submanifold, $y=0$, which divides the entire phase space into two disjoint sectors containing solutions with positive and negative Ricci scalar. This fact will be useful in Section \ref{sec:comp_dyn_sys}, where we will present the compact dynamical system formulation.

At this point, let us discuss the limitations of the standard formulation presented in this section. As we have stressed before, the variable choice \eqref{nc_dyn_variable} works only as long as $F\neq0,\,H\neq0$. $F\ne0$ is a reasonable requirement because otherwise there would be no gravity. The requirement $H\neq0$, however, stops this formalism from any cosmological scenario that has a bounce or a recollapse. Moreover, note that the phase space time variable $\tau=\epsilon\ln a$ is defined such that $\dot\tau =|H|$ rather than $\dot\tau = H$, so that the formulation remains applicable in both expanding and contracting universes \footnote{Any redefinition of time should be such that the new time variable is a monotonically increasing function of time}. The function $\tau(t)$ is not smooth. Although it can be made $\mathcal{C}^1$ if the scale factor at bounce is defined to be unity, it is not $\mathcal{C}^2$ because $\ddot\tau$ involves the derivative of $|H(t)|$ with respect to $t$, which is not
continuous at $H=0$. This flaw is one of the big motivations in formulating the compact dynamical system formulation presented in Section \ref{sec:comp_dyn_sys}.

\section{Some words about cosmography}\label{sec:cosmography}

The basis of the compact dynamical system formulation that we will present in this paper is the use of cosmographic parameters. These are kinematic parameters arising from the Taylor expansion of the Hubble parameter. The $0$-th to $5$-th order cosmographic parameters are as follows:
\begin{subequations}\label{CP}
    \begin{eqnarray}
        H &\equiv& \frac{\dot a}{a}\,,
        \\
        q &\equiv& -\frac{1}{aH^2}\frac{d^2 a}{dt^2} = -1-\frac{\dot H}{H^2} \,,
        \\
        j &\equiv& \frac{1}{aH^3}\frac{d^3 a}{dt^3} = \frac{\ddot H}{H^3} - 3q - 2 \,,
        \\
        s &\equiv& \frac{1}{aH^4}\frac{d^4 a}{dt^4} = \frac{\dddot H}{H^4} + 4j + 3q(q+4) + 6 \,,
        \\
        l &\equiv& \frac{1}{aH^5}\frac{d^5 a}{dt^5} = \frac{H^{(4)}}{H^5} + 5s - 10(j+3q)(q+2) - 24 \,,
    \end{eqnarray}
\end{subequations}
and are called the Hubble, deceleration, jerk, snap, and lerk parameters, respectively. Of course, one could go on to construct even higher-order cosmographic parameters. There is an infinite hierarchy of them \footnote{For some nice reviews see \cite{Dunsby:2015ers, Bolotin:2018xtq}}. As we will see, cosmography in $f(R)$ gravity does not require us to go beyond the lerk parameter (see also \cite{Capozziello:2008qc}). The following relations between the cosmographic parameters are helpful \cite{Dunsby:2015ers,Bolotin:2018xtq}
\begin{subequations}\label{CP_rel}
    \begin{eqnarray}
        j &=& 2q^{2} + q - \frac{dq}{Hdt}\,,
        \\
        s &=& \frac{dj}{Hdt} - j(2 + 3q)\,,
        \\
        l &=& \frac{ds}{Hdt} - s(3+4q)\,.
    \end{eqnarray}
\end{subequations}

A given cosmic evolution, if believed to be the solution of either GR or some $f(R)$ theory, can always be specified by an algebraic relation between a finite number of cosmographic parameters. Let us elaborate on this point using a simple example from GR. The cosmic evolution corresponding to the standard $\Lambda$CDM model of cosmology is a solution of GR plus a cosmological constant, and satisfies the field equation
\begin{equation}\label{eq:lcdm}
    H^2 + \frac{k}{a^2} = \frac{\kappa\rho_0}{3a^3} + \frac{\Lambda}{3}\,.
\end{equation}
The above equation contains $\dot{H}$ and the quantities $\left(\frac{k}{a^2},\frac{\kappa\rho_0}{a^3},\Lambda\right)$. The first and second time derivatives of Eq.\eqref{eq:lcdm} give
\begin{subequations}
    \begin{eqnarray}
        \dot H - \frac{k}{a^2} &=& - \frac{\kappa\rho_0}{2a^3}\,,
        \\
        \frac{\ddot H}{H} + \frac{2k}{a^2} &=& \frac{3}{2}\frac{\kappa\rho_0}{a^3}\,.
    \end{eqnarray}
\end{subequations}
The above equations, along with Eq.\eqref{eq:lcdm}, can be used to solve $\left(\frac{k}{a^2},\frac{\kappa\rho_0}{a^3},\Lambda\right)$ in terms of $(H,\dot{H},\ddot{H})$ from these three equations. Take, now, the third time derivative of Eq.\eqref{eq:lcdm} and substitute the expressions for $\left(\frac{k}{a^2},\frac{\kappa\rho_0}{a^3},\Lambda\right)$. The resulting equation can be expressed entirely in terms of the cosmographic parameters \cite{Dunajski:2008tg}
\begin{equation}
    s + 2(q+j) + qj = 1
\end{equation}
For the special case of spatially flat \emph{or} vacuum cosmologies, one need not go up to the third derivative of Eq.\eqref{eq:lcdm} and the snap parameter is therefore unnecessary. This is the case for spatially flat $\Lambda$CDM cosmologies which gives rise to the simpler and well-known cosmographic condition $j=1$.

A similar exercise could be performed for cosmic solutions of $f(R)$ theories. If a cosmic evolution $a(t)$ is believed to be a solution of some $f(R)$ theory in the presence of a perfect fluid with a constant equation of state $w$, then it must satisfy the modified Friedmann equation \eqref{eq:fried}, which contains up to the second derivative of the Hubble parameter, $\ddot{H}$, and the quantities $\left(\frac{k}{a^2},\frac{\kappa\rho_0}{a^{3(1+w)}}\right)$. Taking a time derivative of this equation gives the modified Raychaudhuri equation \eqref{eq:Raychoudhuri}, which contains up to the third derivative of the Hubble parameter $\dddot{H}$, and the quantities $\left(\frac{k}{a^2},\frac{\kappa\rho_0}{a^{3(1+w)}}\right)$. These two equations can be used to solve the two constants $\left(\frac{k}{a^2},\frac{\kappa\rho_0}{a^{3(1+w)}}\right)$ in terms of the $(H,\dot{H},\ddot{H},\dddot{H})$. Take, now the third derivative of the modified Friedmann equation \eqref{eq:fried} and substitute the expressions for $\left(\frac{k}{a^2},\frac{\kappa\rho_0}{a^{3(1+w)}}\right)$. The resulting equation can be expressed purely in terms of the cosmographic parameters $H,\,q,\,j,\,s,\,l$. One does not need to go beyond the lerk parameter. Again, for spatially flat \emph{or} vacuum cosmological solutions of $f(R)$ theories, one need not go up to the third derivative of Eq.\eqref{eq:fried} and the lerk parameter is unnecessary. For spatially flat \emph{and} vacuum cosmological solutions, one need not even consider the modified Raychudhuri equation Eq.\eqref{eq:Raychoudhuri} and the modified Friedmann equation itself represents a cosmographic relation going only up to the jerk parameter. 

\section{A \textit{model-independent} dynamical system formulation}\label{sec:mod_ind_dyn_sys}

Utilizing the cosmographic parameters, a \emph{model-independent} dynamical system formulation was proposed in Ref.\cite{Chakraborty:2021jku}, which does not require one to specify a-priori an $f(R)$ model, i.e., a functional form of $f(R)$, while still allowing one to obtain a closed autonomous system. In terms of the same dynamical variables, as defined in Eq. \eqref{nc_dyn_variable}, the Friedmann equation and the definition of the Ricci scalar provide two constraints
\begin{subequations}\label{nc_constaints}
    \begin{eqnarray}
        y &=& 1-q+K, \label{eq:nc_2}\\
        z &=& \Omega-x-q \label{eq:nc_3}
    \end{eqnarray}
\end{subequations}
The resulting dynamical system for barotropic perfect fluid of equation of state parameter $\omega$ is
\begin{subequations}\label{dyn_sys_eq}
    \begin{eqnarray}
        \frac{dx}{\epsilon d\tau} &=& x(q-x)+2(x+K+q)-3\Omega (1+\omega)+2\,,\label{eq:nc_ds_x} \\
        \frac{d\Omega}{\epsilon d\tau} &=& -\Omega (x-2q+1+3\omega)\,, \label{eq:nc_ds_Omega}\\
        \frac{dK}{\epsilon d\tau} &=& 2qK\,, \label{eq:nc_ds_K}\\
        \frac{dq}{\epsilon d\tau} &=& 2q^{2}+q-j\,,\label{eq:nc_ds_q}\\
        \frac{dj}{\epsilon d\tau} &=& j(2+3q)+s\,,\label{eq:nc_ds_j}\\
        \frac{ds}{\epsilon d\tau} &=& s(3+4q)+l\,.\label{eq:nc_ds_s}
    \end{eqnarray}   
\end{subequations}
where $y,\,z$ have been eliminated by the constraints \eqref{nc_constaints}. In principle, one could go on and obtain an infinite number of equations since there is an infinite hierarchy of cosmographic parameters. But, as we discussed in section \ref{sec:cosmography}, any cosmic evolution that is a solution of some $f(R)$ gravity can be expressed as an algebraic constraint involving up to the lerk parameter. Therefore one does not need to go beyond Eq.\eqref{eq:nc_ds_s}. When one tries to reconstruct the $f(R)$ starting from a given cosmological solution $a(t)$, one of course believes that there is an underlying $f(R)$ theory of which $a(t)$ is a solution. The dynamical system formulation based on the cosmographic parameters is in some sense an alternative to the reconstruction method. Therefore it is completely justified to take into consideration only up to Eq.\eqref{eq:nc_ds_s}. This particular formulation allows one to study the phase space of all those $f(R)$ theories that can reproduce a cosmic evolution given by the cosmographic condition $l=l(q,j)$, without explicitly reconstructing the underlying $f(R)$.

The dynamical system \eqref{dyn_sys_eq} reveals the existence of another invariant submanifold $K=0$, containing spatially flat solutions. The existence of this invariant submanifold was not apparent from the dynamical system \eqref{nc_dy_sys} because we chose to eliminate the variable $K$ using the Friedmann constraint. This submanifold divides the entire phase space into two disjoint regions, containing spatially positively curved and spatially negatively curved solutions. Present-day observations marginally disfavor spatially negatively curved solutions (see, for example, \cite{DiValentino:2019qzk}), so we can safely assume that the physically interesting cosmological dynamics take place in the $K\geq0$ region. 

For a cosmological solution to be physically realistic, it should be close to the present-day observable universe at $z=0$ ($z$ being the redshift). At $z=0$, $K\gtrsim0,\,q\approx-0.55$, giving a positive value for $y$. Since $y<0$ solutions can never cross into the $y>0$ region and vice versa, we can safely assume that the physically viable cosmological solutions lie in the region $y>0$. Even for non-singular bouncing cosmologies, which we will investigate later, $q\rightarrow-\infty$ at the bounce gives $y>0$. From the above discussions, we can safely say that the physically interesting cosmological dynamics happen within the disjoint region $y\geq0,\,K\geq0$. This fact will be useful in the formulation of the compact phase space in the next section.

It can be noted that 
\begin{equation}\label{x_y_gamma}
    xy\Gamma = 6(x+z-\Omega)+j+3q-2y     
\end{equation}
Eliminating $y$ and $z$ on the right-hand side by using the constraint \eqref{eq:nc_2} and \eqref{eq:nc_3} and using the definition of $\Gamma$ one can write
\begin{equation}
    \frac{1}{y\Gamma}=\frac{6F'H^{2}}{F}=\frac{x}{j-q-2K-2} \;.  
\end{equation}
We recall that the absence of ghost and tachyonic instability requires $F>0,\,F'\geq0$. Assuming that the condition $F>0$ is met, demanding $F'\geq0$ puts the following constraints on the phase space
\begin{equation}\label{v_cond}
    \frac{x}{j-q-2K-2}\geq0\,.    
\end{equation}
Therefore, not the entire phase space is physically viable. The physically viable region of the phase space, where the theory is free from ghost and tachyonic instability, must \emph{necessarily} satisfy the constraint \eqref{v_cond}. One must be careful, though; this is \emph{not} a sufficient condition. It is certainly possible that a region of the phase space satisfying the condition \eqref{v_cond} is plagued by both ghost and tachyonic instability ($F<0$ and $F'<0$). To be sure, one must explicitly calculate $F,\,F'$ along a phase trajectory of interest. In a region of the phase space, if the condition \eqref{v_cond} is satisfied and also the dynamical variable $\Omega\equiv\frac{\kappa \rho}{3FH^2}\geq0$, then one can definitely say that the region is physically viable. We will see that this will be important in example III of section \ref{sec:applications}.
\section{Compact Dynamical System Formulation}\label{sec:comp_dyn_sys}
In this section, following \cite{Goheer:2007wx,Abdelwahab:2011dk,Kandhai:2015pyr,MacDevette:2022hts,Chakraborty:2022}, we present a compact dynamical systems formulation for $f(R)$ gravity. In this approach, all the dimensionless dynamical variables are by definition compact, giving rise to a compact phase space. Such a compact phase space formulation allows us to obtain a global picture of the dynamics, allowing us to investigate fixed points at infinity. Moreover, the formulation is made in such a way that it can be used to investigate the phenomena of non-singular bounce and re-collapse from the phase space point of view. 

The compact phase space formulation presented here is valid for solutions with non-negative global spatial curvature ($k\geq0$) and non-negative Ricci scalar ($R\geq0$). This does not pose any limitation from the physical point of view because, as we have already discussed in the previous sections, this is a disjoint region in the phase space where all the physically interesting cosmological dynamics take place. Two other crucial assumptions that are required to make this formulation work are $f\geq 0$ and the ghost instability never arises ($F>0$). In light of the above requirements, this formulation is very suitable to investigate the phase space of theories like $R+\alpha R^n$ ($\alpha>0,\,n\in\mathbb{R}$) \cite{Abdelwahab:2011dk} or the Hu-Sawicki model \cite{MacDevette:2022hts}.
 
The compact phase space formulation starts with rewriting the Friedmann equation as
\begin{equation}
\left(3H+\frac{3}{2}\frac{\dot{F}}{F}\right)^{2}+\frac{3}{2}\left(\frac{f}{F}+\frac{6k}{a^{2}}\right)=\frac{3\rho}{F}+\frac{3R}{2}+\left(\frac{3}{2}\frac{\dot{F}}{F}\right)^{2}\,,  \label{com_ds_friedmann_eq}    
\end{equation}
allowing for the introduction of a new normalization factor $D$ defined as
\begin{equation}\label{D}
    D^{2}=\left(3H+\frac{3}{2}\frac{\dot{F}}{F}\right)^{2}+\frac{3}{2}\left(\frac{f}{F}+\frac{6k}{a^{2}}\right)\,.
\end{equation}
Note that, for $k>0$, the quantity $D$ is positive definite $D>0$. A new set of dynamical variables are then defined as,
\begin{equation}\label{comp_dyn_var}
    \Bar{x}=\frac{3}{2}\frac{\dot{F}}{F}\frac{1}{D}, ~~~ 
    \Bar{y}=\frac{3}{2}\frac{R}{D^{2}}, ~~~
    \Bar{z}=\frac{3}{2}\frac{f}{F}\frac{1}{D^{2}}, ~~~
    \Bar{Q}=\frac{3H}{D}, ~~~ 
    \Bar{\Omega}=\frac{3\rho}{F}\frac{1}{D^{2}}, ~~~
    \Bar{K}=\frac{9k}{a^{2}}\frac{1}{D^{2}}.
\end{equation}
We now have two independent constraints equations, one derived from the Friedmann equation \eqref{com_ds_friedmann_eq} and the other just from the definition of $D$ in \eqref{D},
\begin{subequations}\label{comp_constr}
    \begin{eqnarray}
        \Bar{\Omega} + \Bar{y} + \Bar{x}^{2} &=& 1\,,\label{comp_constr_1}\\
        (\Bar{Q} + \Bar{x})^{2} + \Bar{z} + \Bar{K} &=& 1\,.\label{comp_constr_2}
    \end{eqnarray}
\end{subequations}
The above constraints, together with the assumptions $k\geq0,\,R\geq0,\,f\geq0,\,F>0$, compactify the phase space
\begin{eqnarray}
    \Bar{x}\in [-1,1], ~~~~~~~~\Bar{Q}\in [-2,2] ~~~~~~~\Bar{y}, ~\Bar{z}, ~\Bar{\Omega}, ~\Bar{K} \in [0,1] \nonumber \\
\end{eqnarray}
A new phase space-time variable $\Bar{\tau}$ is also defined as
\begin{equation}
    d\Bar{\tau} = D dt\,.    
\end{equation}
Note that the normalisation $D$ is chosen in such a way so that it may be strictly positive in order for the function $\Bar{\tau}(t)$ to be monotonically increasing\footnote{One can argue what happens if $D^2=0$. This is possible only for the very special situation $3H+\frac{3}{2}\frac{\dot{F}}{F} = f = k = 0$. However, this is not possible identically. For example, even if one considers a spatially flat universe $k=0$, note that $f(R)$ cannot be identically zero, otherwise there would be no gravity whatsoever. $f(R)=0$ corresponds to $\bar{z}=0$, and this is not an invariant submanifold. Similarly, one cannot have $3H+\frac{3}{2}\frac{\dot{F}}{F}=0$ identically, as this would mean $\bar{Q}+\bar{x}=0$ and this, again, is not an invariant submanifold. Since $D^2$ cannot identically vanish, one can still proceed with the dynamical system formulation}. The compact and the usual non-compact dynamical variables and time variables are related as follows
\begin{equation}\label{mapping}
    x=\frac{2\Bar{x}}{\Bar{Q}}, ~~
    y=\frac{\Bar{y}}{\Bar{Q}^{2}}, ~~
    z=\frac{\Bar{z}}{\Bar{Q}^{2}}, ~~
    K=\frac{\Bar{K}}{\Bar{Q}^{2}}, ~~
    \Omega =\frac{\Bar{\Omega}}{\Bar{Q}^{2}}, ~~
    \epsilon d\tau = H dt = \frac{\bar{Q}}{3}d\Bar{\tau}.   
\end{equation}
The case of $\Bar{Q}=0$ and hence $H=0$ (point which connects expanding and contracting trajectories) is allowed within the compact formalism but not within the non-compact formalism. This was discussed in Section \ref{sec:dyn_sys} but also is apparent from the relationships above.  Using the constraint equations, we can eliminate $\bar{z},\,\bar{\Omega}$ to arrive at the following autonomous system

\begin{subequations}\label{eq:compactdynsys}
\begin{eqnarray}
\frac{d\bar{x}}{d\bar{\tau}} =&& \frac{1}{6}\Big[-2 \bar{x}^{2} \bar{y} \Gamma+(1-3 \omega)(1-\bar{x}^2-\bar{y})+2 \bar{y}+4\left(\bar{x}^{2}-1\right)\left(1-\bar{Q}^{2}-\bar{x} \bar{Q}\right)+\bar{x}(\bar{Q}+\bar{x})\Big]\\
       \frac{d\bar{y}}{d\bar{\tau}} =&& -\frac{1}{3}\bar{y}\Big[ (\bar{Q}+\bar{x})\Big(2\bar{y} -(1+3\omega)(1-\bar{x}^2-\bar{y})   +4\bar{x}\bar{Q}   \Big)-2\bar{Q} -4\bar{x}   +2 \bar{x} \Gamma(\bar{y}-1) +4\bar{x}\bar{K} \nonumber \\
       &&\left((1+3 \omega)(1-\bar{x}^2-\bar{y})-2 \bar{y}\right)+4\bar{K}(1-\bar{x}^2)\Big]\,,\\
       \frac{d\bar{Q}}{d\bar{\tau}}= &&\frac{1}{6}\Big[-4\bar{x}\bar{Q}^3+\bar{x}\bar{Q}(5+3\omega)(1-\bar{x}\bar{Q})-\bar{Q}^2(1-3\omega)-\bar{Q}\bar{x}^3(1+3\omega)-3\bar{y}\bar{Q}(1+\omega)(\bar{Q}+\bar{x})\nonumber\\
       &&+2\bar{y}(1-\Gamma \bar{Q}\bar{x}) - 2\bar{K}(1+2\bar{x}\bar{Q})\Big]\,,\\
       \frac{d\bar{K}}{d\bar{\tau}}=&&-\frac{1}{3} \bar{K}\Big[ (\bar{Q}+\bar{x})\Big( -(1+3\omega)(1-\bar{x}^2-\bar{y})  +4\bar{x}\bar{Q}   +2\bar{y}\Big)+4\bar{x}(\bar{K}-1)+2 \bar{x}\bar{y} \Gamma \Big]\,,
\end{eqnarray}
\end{subequations}

where $\Gamma\equiv\frac{F}{RF'}$ is the same auxiliary quantity as defined in Eq.\eqref{Gamma}. Just like the usual dynamical system formulation of Section \ref{sec:dyn_sys}, the applicability of this compact formulation is also limited by the requirement that it should be possible to express $\Gamma$ in terms of $\bar{y}$ and $\bar{z}$, thus closing the system \eqref{eq:compactdynsys}.

\section{A \textit{model-independent} compact dynamical system formulation}\label{sec:mod_ind_comp_dyn_sys}
Inspired by the model-independent dynamical system formulation presented in Section \ref{sec:mod_ind_dyn_sys}, we now try to construct a similar model-independent formulation, but in terms of the compact dynamical variables. Such a formulation will allow us to investigate the global phase space structure of those $f(R)$ theories that satisfy a particular cosmographic constraint without requiring us to reconstruct their exact mathematical form, which might be too complicated for any practical purpose. The only restriction on the mathematical form of the $f(R)$ theories are the ones required for the compact formulation to work, namely $f\geq0,\,F>0$. We will again make use of the cosmographic deceleration and jerk parameter.

In constructing the compact dynamical system \eqref{eq:compactdynsys}, we chose to eliminate $\bar{\Omega},\,\bar{z}$. If we instead choose to eliminate $\bar{y},\,\bar{z}$, then we arrive at the following dynamical system:
\begin{subequations}\label{eq:comp_dyn_sys_1}
    \begin{eqnarray}
        \frac{d\Bar{x}}{d\Bar{\tau}}&=&\frac{1}{6}  \Big[2\left(-1+2\Bar{K}+\Bar{x}\Bar{Q}-2\Bar{x}^{2}\Bar{K}+2\Bar{Q}^{2}-2\Bar{x}^{2}\Bar{Q}^{2}-\Bar{x}^{3}\Bar{Q}+\Bar{x}^{4}\right)-\Bar{\Omega}(1+3\omega)+3\Bar{x}\Bar{\Omega}(1+\omega)(\Bar{Q}+\Bar{x}) \nonumber \\
        &&-2\Bar{x}^{2}\left(1-\Bar{\Omega} -\Bar{x}^{2}\right)\Gamma \Big], \\
        \frac{d\Bar{\Omega}}{d\Bar{\tau}}&=&\frac{\Bar{\Omega}}{3}\Big[3(\Bar{x}\Bar{\Omega}+\Bar{Q}\Bar{\Omega}-\Bar{Q})(1+\omega)-2\Bar{x}\left(1-\Bar{\Omega} -\Bar{x}^{2}\right)\Gamma-4\Bar{x}\Bar{Q}^{2}-4\Bar{x}\Bar{K}-2\Bar{x}^{2}\Bar{Q} +2\Bar{x}^{3}\Big], \\
        \frac{d\Bar{Q}}{d\Bar{\tau}}&=&\frac{1}{6} \Big[3\Bar{Q}\Bar{\Omega}(1+\omega)(\Bar{Q}+\Bar{x})-2\Bar{K}-4\Bar{Q}^{2}+2\Bar{x}\Bar{Q}-2(\Bar{Q}\Bar{x}\Gamma-1)\left(1-\Bar{\Omega} -\Bar{x}^{2}\right) \nonumber \\      &&-4\Bar{x}\Bar{Q}\Bar{K}-2\Bar{x}^{2}\Bar{Q}^{2}+2\Bar{x}^{3}\Bar{Q} -4\Bar{x}\Bar{Q}^{3}\Big], \\
        \frac{d\Bar{K}}{d\Bar{\tau}}&=&\frac{\Bar{K}}{6} \Big[4\Bar{x}-4\Bar{Q}-8\Bar{x}\Bar{K}-8\Bar{x}\Bar{Q}^{2}-4\Bar{x}^{2}\Bar{Q}+4\Bar{x}^{3}+6\Bar{\Omega}(1+\omega)(\Bar{Q}+\Bar{x})-4\Bar{x}\left(1-\Bar{\Omega} -\Bar{x}^{2}\right)\Gamma\Big].
    \end{eqnarray}  
\end{subequations}
The auxiliary variable $\Gamma$, which encodes the information about the explicit form of $f(R)$, enters the dynamical system only through the combination
\begin{equation}
    \Bar{x}(1-\Bar{\Omega} -\Bar{x}^{2})\Gamma = \bar{x}\bar{y}\Gamma\,.
\end{equation}
Using the mapping \eqref{mapping}, the relations \eqref{x_y_gamma} and the constraints \eqref{comp_constr}, it is possible to express the above expression in terms of the cosmographic deceleration and jerk parameter 
\begin{equation}\label{eq:cond_a}
   \Bar{x}\left(1-\Bar{\Omega} -\Bar{x}^{2}\right)\Gamma = \frac{\Bar{Q}^{3}}{2}(j-q-2)-\Bar{K}\Bar{Q} \,. 
\end{equation}
Substituting this back into the system \eqref{eq:comp_dyn_sys_1} gives the dynamical system
\begin{subequations}\label{dyn_generic}
    \begin{eqnarray}
        \frac{d\Bar{x}}{d\Bar{\tau}}&=&\frac{1}{6}  \Big[2\left(-1+2\Bar{K}+\Bar{x}\Bar{Q}-2\Bar{x}^{2}\Bar{K}+2\Bar{Q}^{2}-2\Bar{x}^{2}\Bar{Q}^{2}-\Bar{x}^{3}\Bar{Q}+\Bar{x}^{4}\right)-\Bar{\Omega}(1+3\omega)+3\Bar{x}\Bar{\Omega}(1+\omega)(\Bar{Q}+\Bar{x}) \nonumber \\
        &&-\Bar{x}\Bar{Q}^{3}(j-q-2)+2\Bar{x}\Bar{K}\Bar{Q}\Big]\,,
        \\
        \frac{d\Bar{\Omega}}{d\Bar{\tau}}&=&\frac{\Bar{\Omega}}{3}\Big[3(\Bar{x}\Bar{\Omega}+\Bar{Q}\Bar{\Omega}-\Bar{Q})(1+\omega)-\Bar{Q}^{3}(j-q-2)+2\Bar{K}\Bar{Q}-4\Bar{x}\Bar{Q}^{2}-4\Bar{x}\Bar{K}-2\Bar{x}^{2}\Bar{Q} +2\Bar{x}^{3}\Big]\,, 
        \\   
        \frac{d\Bar{Q}}{d\Bar{\tau}}&=&\frac{1}{6} \Big[3\Bar{Q}\Bar{\Omega}(1+\omega)(\Bar{Q}+\Bar{x})-2\Bar{K}-4\Bar{Q}^{2}+2\Bar{x}\Bar{Q}-\Bar{Q}^{4}(j-q-2)+2\Bar{K}\Bar{Q}^{2} +2\left(1-\Bar{\Omega} -\Bar{x}^{2}\right) \nonumber \\ 
        &&-4\Bar{x}\Bar{Q}\Bar{K} -2\Bar{x}^{2}\Bar{Q}^{2} +2\Bar{x}^{3}\Bar{Q}-4\Bar{x}\Bar{Q}^{3}\Big]\,, 
        \\
        \frac{d\Bar{K}}{d\Bar{\tau}}&=&\frac{\Bar{K}}{6} \Big[4\Bar{x}-4\Bar{Q}-8\Bar{x}\Bar{K}-8\Bar{x}\Bar{Q}^{2}-4\Bar{x}^{2}\Bar{Q}+4\Bar{x}^{3}+6\Bar{\Omega}(1+\omega)(\Bar{Q}+\Bar{x})-2\Bar{Q}^{3}(j-q-2)+4\Bar{K}\Bar{Q}\Big]\,,
    \end{eqnarray}    
\end{subequations}
which contains no $\Gamma$-term, but instead contains the cosmographic parameters $q,\,j$. The last piece of the trick is to use the definition of the Ricci scalar in Eq.\eqref{eq:nc_2}, the mapping \eqref{mapping} and the Friedmann constraint \eqref{comp_constr_1} in such a way as to be able to write 
\begin{eqnarray}\label{eq:dec_non_vacuum}
q = 1 - \frac{\bar{y} - \bar{K}}{\bar{Q}^2} = 1-\frac{1-\Bar{\Omega} -\Bar{x}^{2}-\Bar{K}}{\Bar{Q}^{2}}\,.    
\end{eqnarray}
If a cosmic evolution can be specified by an algebraic relation of the form $j=j(q)$, then the dynamical system \eqref{dyn_generic}, together with the relation \eqref{eq:dec_non_vacuum}, provides a model-independent dynamical system, in the sense that it can be used to investigate the phase space of all such $f(R)$ theories that support the cosmology $j=j(q)$.

We notice the existence of two invariant submanifolds corresponding to the spatially flat case ($\bar{K}=0$) and the vacuum case ($\bar{\Omega}=0$). At the intersection of these two invariant submanifolds, which is an invariant submanifold itself corresponding to the spatially flat vacuum case, the flow is 2-dimensional.

The physically viable region of the phase space, where the theory is free from ghost and tachyonic instability, is constrained by the conditions $F>0,\,F'\geq0$. Noting that 
\begin{equation}
    \frac{2}{3}\frac{F' D^{2}}{F}=\frac{1}{\left(1-\Bar{\Omega} -\Bar{x}^{2}\right)\Gamma}=\frac{2\Bar{x}}{\Bar{Q}^{3}(j-q-2)-2\Bar{K}\Bar{Q}}\label{eq:cond_b}\,,
\end{equation}
the physically viable region of the phase space must necessarily satisfy the condition
\begin{equation}\label{eq:cond_c}
    \frac{2\Bar{x}}{\Bar{Q}^{3}(j-q-2)-2\Bar{K}\Bar{Q}}\geq0\,.
\end{equation}

The above condition is equivalent to the condition \eqref{v_cond}. We recall that the condition \eqref{v_cond} was only a necessary condition but not a sufficient condition for the physical viability of a region of the phase space. This is because the condition \eqref{v_cond} is also satisfied when both ghost and tachyonic instability are present simultaneously. However, we recall that $F>0$ was actually a key assumption to construct the compact phase space. Without this assumption, the dynamical variables defined in \eqref{comp_dyn_var} would not be bounded. Therefore, the condition for the absence of ghost instability, namely $F>0$, is already implicitly present in the compact phase space analysis. Thus, for a compact phase space analysis, we can safely take the condition \eqref{eq:cond_c} to single out the physically viable regions of phase space.

\section{Applications} \label{sec:applications}
To illustrate the applicability of the model-independent compact dynamical system formulation constructed in the previous section, we now apply the formulation to three simple cosmological scenarios: two nonsingular bounce scenarios and a cyclic cosmology. Here, by ``simple'', we mean that, although the cosmographic relation corresponding to a very generic cosmological solution of an $f(R)$ theory can involve up to the lerk parameter, the particular examples we consider here correspond to a much simpler cosmographic constraint $j=j(q)$. To do this, we proceed as follows: Firstly, we specify a cosmological scenario as a given function of time, $a(t)$ or $H(t)$. From there, it is straightforward to calculate the deceleration parameter
\begin{equation}\label{eq:decel}
    q(t) = - \frac{1}{aH^2}\frac{d^2 a}{dt^2} = - 1 - \frac{\dot H}{H^2}\,,
\end{equation}
and the jerk parameter 
\begin{equation}\label{eq:jerk}
    j(t) = \frac{1}{aH^3}\frac{d^3 a}{dt^3} = - 2 - 3q + \frac{\ddot H}{H^3}\,.
\end{equation}
Provided that we can use the two equations above to write $j=j(q)$ \footnote{It does not necessarily require $q=q(t)$ to be invertible. In all the three subsequent examples considered, $q(t)$ is an even function and therefore non-invertible. Nonetheless, we are able to write $j=j(q)$.}, we can now use the expression \eqref{eq:dec_non_vacuum} for $q$ to close the dynamical system. In practice, we note that $q$ and $j$ appear in the system only through the combination $Q^3 (j-q-2)$, so for each of the cosmological scenarios, we will present the expression of this quantity explicitly.

The cosmic evolution corresponding to the fixed points can be obtained by calculating the deceleration parameter $q=-1-\frac{\dot H}{H^2}$ using the expression \eqref{eq:dec_non_vacuum},  except for those fixed points for which the limiting value of the quantity $\frac{\dot{H}}{H^2}$ on approach to the fixed point is indeterminate; $\frac{\dot{H}}{H^2}\rightarrow\pm\frac{0}{0}$. This occurs when
\begin{equation}\label{eq:dec_undef}
Q=0\,\,,\,\,  \Bar{\Omega} + \Bar{x}^{2} + \Bar{K} = 1\,.  
\end{equation}
For such points, we cannot determine the cosmic evolution by this method. Let us elaborate on this point from a more geometric point of view. Eq.\eqref{eq:dec_non_vacuum} can be written as
\begin{eqnarray}
\Bar{\Omega} + \Bar{x}^{2} + \Bar{K} + (1-q)\Bar{Q}^{2} = 1\,,
\end{eqnarray}
which shows that hypersurfaces of constant $q$ correspond in general to a family of 3-dimensional hypersurfaces within the 4-dimensional phase space. All these hypersurfaces of constant $q$ intersect at the curve given by Eq.\eqref{eq:dec_undef}. If a fixed point lies on one specific such hypersurface, then its cosmology can be determined unambiguously. However, if a fixed point lies at the intersection of all these hypersurfaces, i.e., on the curve \eqref{eq:dec_undef}, then one cannot uniquely identify the cosmology corresponding to such a fixed point.\footnote{One can think of this situation as being similar to what one encounters is plane polar coordinates $(r,\theta)$. The origin does not have a unique $\theta$ coordinate as all $\theta=constant$ lines intersect at the origin.} This is an artefact of the particular choice of the compact variables \eqref{comp_dyn_var}.\footnote{One may not encounter this issue if one uses, say, the Poincar\'e compactification prescription \cite{Abdelwahab:2007jp}. The particular compactification prescription used here is used because of its advantage in the clear depiction of bouncing trajectories.} 
As we will see below, all three examples we consider have fixed points like this, so we need to be careful while interpreting the cosmology corresponding to such points.

We should emphasize that the success of the formulation that we develop here depends crucially on being able to write a cosmographic relation $j=j(q)$. This is the case for the examples we considered here, but this will not be the case in general, since the cosmographic condition corresponding to a generic cosmological solution of an $f(R)$ theory will also contain snap and lerk parameter.
\subsection{Example-I}
As our first example, let us take the ansatz
\begin{equation}\label{eq:model-I}
    H(t)=\frac{\beta t}{\left(\alpha^{2}+t^{2}\right)}\,,    
\end{equation}
Where $-\infty <t<\infty $, $\alpha$ being the characteristic time-scale, i.e., $[\alpha]=[t]$ and $\beta>0$ is the dimensionless (stretching) constant. For $|t|\gg \alpha$,  $H(t)\approx \beta/t$ and for $|t|\ll \alpha$, $H(t)\approx \beta t/\alpha^{2}$. This ansatz therefore represents a cosmic evolution that connects a power law contraction phase $a(t)\sim(-t)^\beta$ to a power law expanding phase $a(t)\sim t^\beta$ through a nonsingular bounce of the form $a(t)\sim e^{\beta t^2 / (2\alpha^2)}$. This ansatz was considered in the Ref.\cite{Arora:2022dti} where the authors were concerned about the behaviour of small shear during such an evolution. Following \cite{Arora:2022dti}, we consider the numerical value of $\beta$ to be the same as $\alpha$ for the sake of simplicity.

\begin{figure}[H]
	\begin{center}
		\includegraphics[width=7cm]{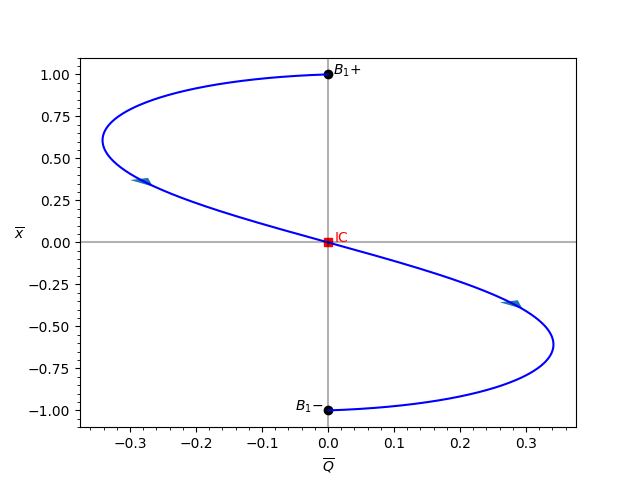}
		\includegraphics[width=7cm]{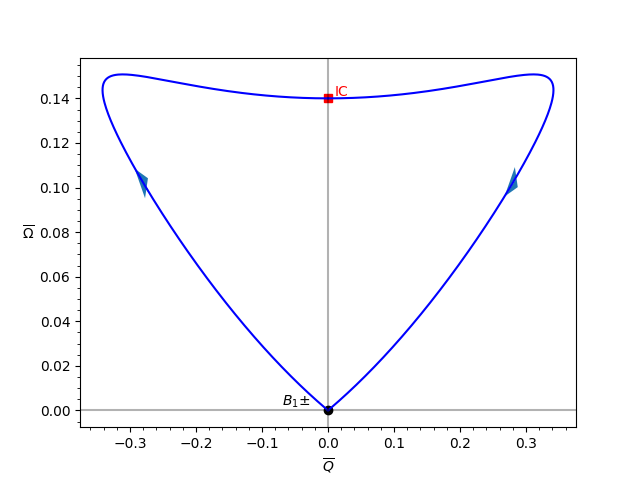}
		\includegraphics[width=7cm]{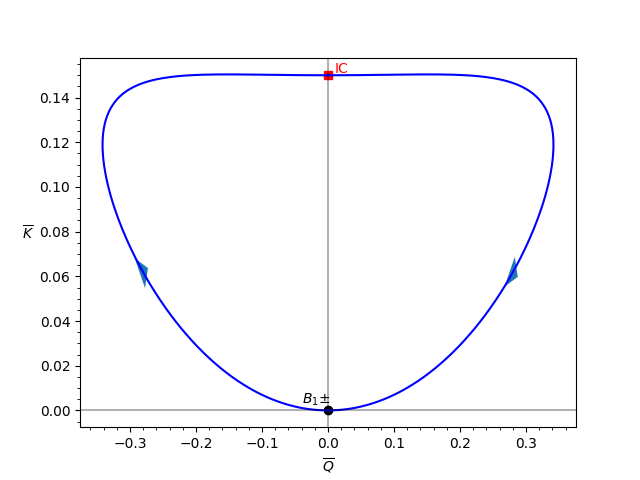}
            \includegraphics[width=7cm]{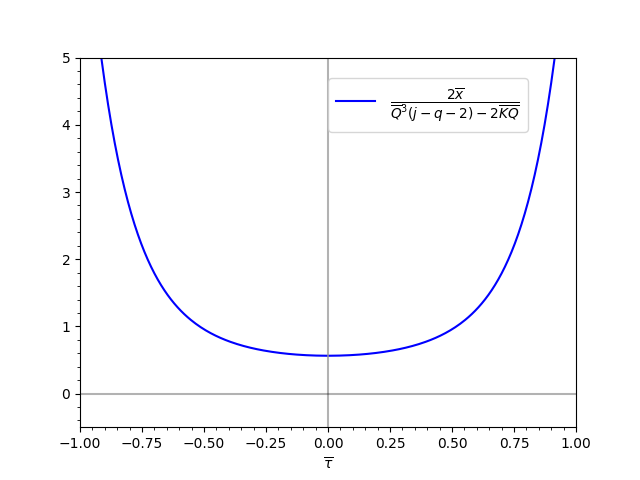}
	\end{center}
	\caption{Parametric plots $\bar{x}(\bar{\tau})$ v/s $\bar{Q}(\bar{\tau})$ (upper left), $\bar{\Omega}(\bar{\tau})$ v/s $\bar{Q}(\bar\tau)$ (upper right) and $\bar{K}(\bar\tau)$ v/s $\bar{Q}(\bar\tau)$ (lower left) for a trajectory passing through the point $(\bar{x},\bar{\Omega},\bar{Q},\bar{K})=(0,0.14,0,0.15)$ for the ansatz \eqref{eq:model-I} with $\omega=0,\,\alpha=2/3$. The lower right panel shows the quantity $\frac{2\Bar{x}}{\Bar{Q}^{3}(j-q-2)-2\Bar{K}\Bar{Q}}$ remains positive throughout the entire evolution history of the bouncing cosmology, signalling a stable bouncing solution.}
\label{fig:model_I_a}
\end{figure}

The deceleration parameters can be calculated from Eq.\eqref{eq:decel}
\begin{equation}
    q = -1-\frac{\alpha^{2}-t^{2}}{\alpha t^{2}}\,,\label{eq:q}
\end{equation}
inverting, which we get
\begin{equation}
    t^{2}=\frac{\alpha^{2}}{1-\alpha(1+q)}\,.
\end{equation}
The jerk parameter can be calculated from Eq.\eqref{eq:jerk}. Using the above expression for $t^2$ in terms of $q$ we get
\begin{equation}
    j = \frac{\alpha -2}{\alpha^{2}}[2-\alpha(2+3q)]\,.\label{eq:j-I}
\end{equation}
Finally, we obtain
\begin{equation}
    \Bar{Q}^{3}(j-q-2)=\frac{2(2\alpha-3)}{\alpha}\Bar{Q}(1-\Bar{\Omega}-\Bar{x}^{2}-\Bar{K})-\frac{4\Bar{Q}^{3}}{\alpha^{2}}(2\alpha^{2}-3\alpha+1)\,,    
\end{equation}
which gives us the following dynamical system
\begin{subequations}
    \begin{eqnarray}
        \frac{d\Bar{x}}{d\Bar{\tau}}&=&\frac{1}{6}  \Big[2\left(-1+2\Bar{K}+\Bar{x}\Bar{Q}-2\Bar{x}^{2}\Bar{K}+2\Bar{Q}^{2}-2\Bar{x}^{2}\Bar{Q}^{2}-\Bar{x}^{3}\Bar{Q}+\Bar{x}^{4}\right)-\Bar{\Omega}(1+3\omega)+3\Bar{x}\Bar{\Omega}(1+\omega)(\Bar{Q}+\Bar{x}) \nonumber \\
        &&-\frac{2(2\alpha-3)}{\alpha}\Bar{x}\Bar{Q}(1-\Bar{\Omega}-\Bar{x}^{2}-\Bar{K})+\frac{4\Bar{x}\Bar{Q}^{3}}{\alpha^{2}}\left(2\alpha^{2}-3\alpha+1\right)+2\Bar{x}\Bar{K}\Bar{Q}\Big]\,,
        \\
        \frac{d\Bar{\Omega}}{d\Bar{\tau}}&=&\frac{\Bar{\Omega}}{3}\Big[3(\Bar{x}\Bar{\Omega}+\Bar{Q}\Bar{\Omega}-\Bar{Q})(1+\omega)-\frac{2(2\alpha-3)}{\alpha}\Bar{Q}(1-\Bar{\Omega}-\Bar{x}^{2}-\Bar{K})+\frac{4\Bar{Q}^{3}}{\alpha^{2}}\left(2\alpha^{2}-3\alpha+1\right)+2\Bar{K}\Bar{Q} -4\Bar{x}\Bar{Q}^{2}  \nonumber \\
        &&-4\Bar{x}\Bar{K}-2\Bar{x}^{2}\Bar{Q} +2\Bar{x}^{3}\Big], 
        \\ 
        \frac{d\Bar{Q}}{d\Bar{\tau}}&=&\frac{1}{6} \Big[3\Bar{Q}\Bar{\Omega}(1+\omega)(\Bar{Q}+\Bar{x})-2\Bar{K}-4\Bar{Q}^{2}+2\Bar{x}\Bar{Q}-\frac{2(2\alpha-3)}{\alpha}\Bar{Q}^{2}(1-\Bar{\Omega}-\Bar{x}^{2}-\Bar{K})+\frac{4\Bar{Q}^{4}}{\alpha^{2}}\left(2\alpha^{2}-3\alpha+1\right)  \nonumber \\
        &&+2\Bar{K}\Bar{Q}^{2}+2\left(1-\Bar{\Omega}-\Bar{x}^{2}\right) -4\Bar{x}\Bar{Q}\Bar{K}-2\Bar{x}^{2}\Bar{Q}^{2}+2\Bar{x}^{3}\Bar{Q}-4\Bar{x}\Bar{Q}^{3}\Big]\,, 
        \\
        \frac{d\Bar{K}}{d\Bar{\tau}}&=&\frac{\Bar{K}}{6} \Big[4\Bar{x}-4\Bar{Q}-8\Bar{x}\Bar{K}-8\Bar{x}\Bar{Q}^{2}-4\Bar{x}^{2}\Bar{Q}+4\Bar{x}^{3}+6\Bar{\Omega}(1+\omega)(\Bar{Q}+\Bar{x})-\frac{4(2\alpha-3)}{\alpha}\Bar{Q}(1-\Bar{\Omega}-\Bar{x}^{2}-\Bar{K})\nonumber \\ 
        &&+\frac{8\Bar{Q}^{3}}{\alpha^{2}}\left(2\alpha^{2}-3\alpha+1\right)+4\Bar{K}\Bar{Q}\Big]\,.
    \end{eqnarray}  
\end{subequations}
The fixed points of this dynamical system for the case of dust ($\omega=0$) and assuming the parameter value $\alpha=2/3$ are listed in Table \ref{tab:generic_crit_model_I}. 
\begin{table}[H]
\resizebox{\textwidth}{!}{
    \centering
    \begin{tabular}{|c|c|c|c|c|}\hline
 \begin{tabular}{@{}c@{}} Fixed \\ points \end{tabular} & Coordinates $(\Bar{x},\Bar{\Omega},\Bar{Q},\Bar{K})$ & Jacobian eigenvalues & Stability & Cosmology \\ \hline
  $B_{1\pm}$ & $(\pm1,0,0,0)$ & $\big\{\pm\frac{4}{3},\pm\frac{4}{3},\pm\frac{2}{3},\pm\frac{2}{3}\big\}$ & \begin{tabular}{@{}c@{}} $(-)$ attractor \\ $(+)$ repeller \end{tabular} & \begin{tabular}{@{}c@{}} $|t|\rightarrow\infty$ limit of \\ $a(t)\sim|t|^{\frac{2}{3}}$ \end{tabular} \\ \hline
  $B_{2\pm}$ & $(\pm 0.53, 0, \mp 1.20,0)$ & $\big\{\mp 1.70, \pm 1.20, \mp 0.40, \mp 0.35\big\}$ & Saddle & $a(t)\sim|t|^{\frac{2}{3}}$ \\ \hline
  $B_{3\pm}$ & $(\pm 0.923,0,\pm 0.545,0)$ & $\big\{\mp 0.775, \mp 0.615, \mp 0.544, \pm 0.181\big\}$ & Saddle & $a(t)\sim|t|^{\frac{2}{3}}$ \\ \hline
  $B_{4\pm}$ & $\left(0,\frac{2}{3},\pm \sqrt{\frac{2}{3}},0\right)$ & $\left\{\pm \frac{\sqrt{73}+5}{6 \sqrt{6}},\mp \sqrt{\frac{2}{3}},\pm \frac{1}{3}\sqrt{\frac{2}{3}},\mp \frac{\sqrt{73}-5}{6 \sqrt{6}}\right\}$ & Saddle & $a(t)\sim|t|^{\frac{2}{3}}$ \\ 
  \hline
  $B_{5\pm}$ & $\left(\pm \frac{3}{\sqrt{17}},\frac{12}{17},\pm \frac{2}{\sqrt{17}},0\right)$ & $\left\{\mp \frac{2(\sqrt{17}+\sqrt{170})}{51},\pm \frac{8}{3 \sqrt{17}},\pm \frac{2}{\sqrt{17}},\pm \frac{2(\sqrt{170}-\sqrt{17})}{51} \right\}$ & Saddle & $a(t)\sim|t|^{\frac{2}{3}}$ 
  \\ \hline
  $B_{6\pm}$ & $\left(\pm\frac{1}{\sqrt{3}},0,0,\frac{2}{3}\right)$ & $\left\{\mp \frac{4}{3 \sqrt{3}},\mp \frac{2}{3 \sqrt{3}},0,0\right\}$ & \begin{tabular}{@{}c@{}} $(+)$ attractor \\ or saddle \\ $(-)$ repeller \\ or saddle\end{tabular} & Indeterminate 
  \\ \hline  
  $B_{7}$ & $\left(0,\frac{2}{5},0,\frac{3}{5}\right)$ & $\left\{-\sqrt{\frac{2}{15}},\sqrt{\frac{2}{15}},0,0\right\}$ & Saddle & Indeterminate 
  \\ \hline
  \end{tabular}}
    \caption{Fixed points, Jacobian eigenvalues, stability and the respective cosmologies for the cosmic evolution corresponding to the ansatz \eqref{eq:model-I} with $\omega=0$ and $\alpha=2/3$. Cosmic evolution corresponding to the fixed points $B_{1\pm},\,B_{6\pm},\,B_7$ cannot be determined as, according to the expression \eqref{eq:dec_non_vacuum}, $q$ is indeterminate.}
    \label{tab:generic_crit_model_I}
\end{table}
In general, the phase space is 4-dimensional. Figure \ref{fig:model_I_a} shows different projections of a phase trajectory that corresponds to a non-singular bounce. The last panel in the figure proves that throughout its entire evolution history, the bouncing cosmology encounters neither ghost nor tachyonic instabilities.

One can note that the cosmology corresponding to the fixed points $B_{1\pm},\,B_{6\pm},\,B_7$ cannot be determined from the deceleration parameter, but has to be found in other ways. We note that, for all these fixed points, $Q=0$, i.e., $H\equiv\frac{\dot a}{a}=0$. This can imply either a static solution $a(t)=constant$, or the $|t|\rightarrow\infty$ limit of the asymptotic form of the evolution ($a(t)\sim|t|^{2/3}$) where $\dot{a}(t)$ diverges slower than $a(t)$, essentially making $H(t)\rightarrow0$. We identify the cosmology corresponding to the fixed points $B_{1\pm}$ with the latter option, which is supported by the first panel of Figure \ref{fig:model_I_a}. This gives a picture of a bouncing cosmology consistent with the ansatz \eqref{eq:model-I}. Unfortunately, for $B_{6\pm},\,B_7$, we still cannot exactly identify the corresponding cosmology.

\begin{figure}[H]
    \centering
    \includegraphics[width=6cm]{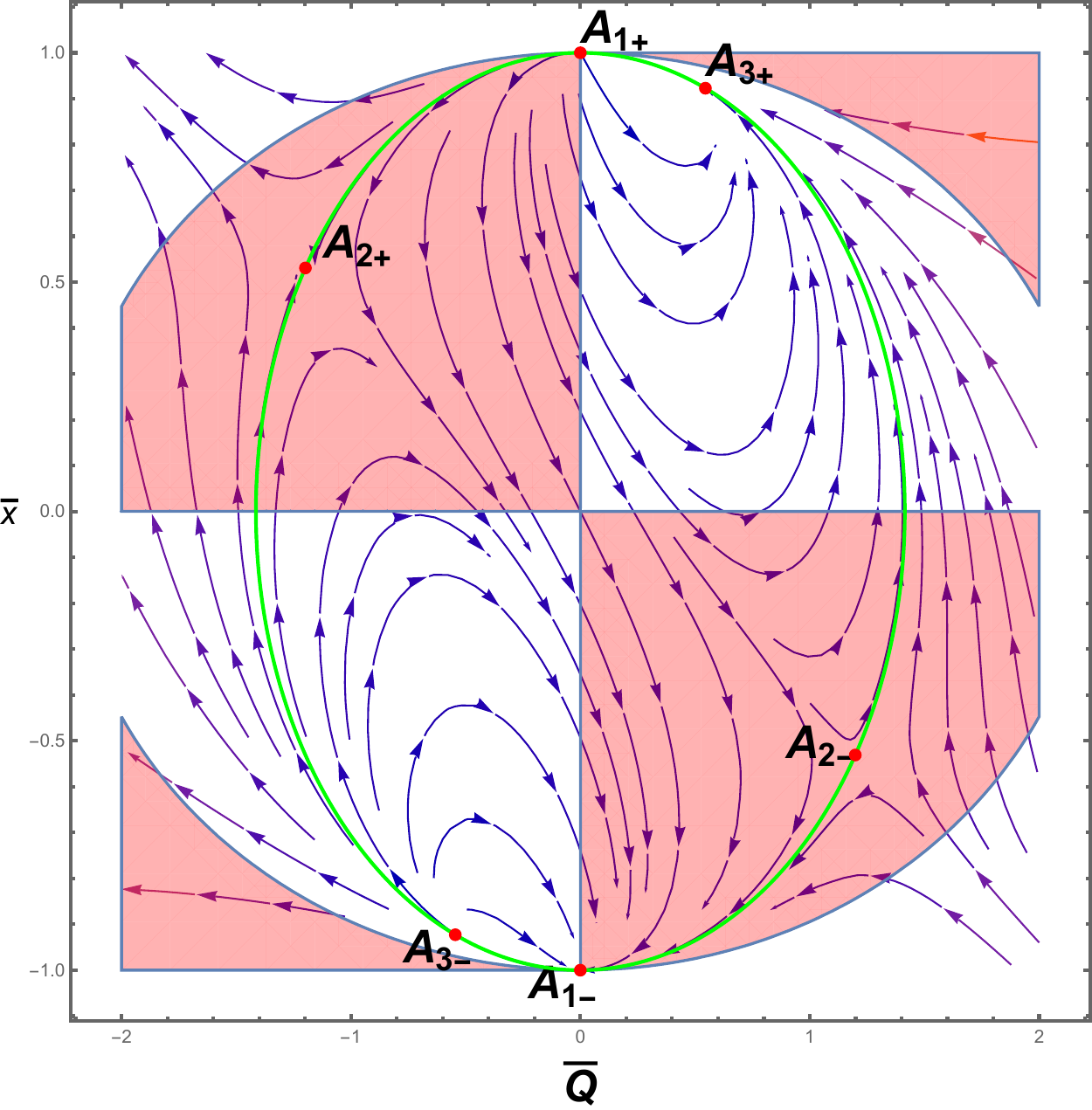}
    \caption{Phase portrait for the spatially flat vacuum case for the ansatz \eqref{eq:model-I} with $\alpha=2/3$. The shaded region is the physically viable region where the $f(R)$ theory is free from ghost and tachyonic instability ($F>0,\,F'>0$). The green ellipse corresponds to the invariant submanifold $q=\frac{1-\alpha}{\alpha}=\frac{1}{2}$.}
    \label{fig:model_I}
\end{figure}

\begin{figure}[H]
\centering
\includegraphics[scale=0.35]{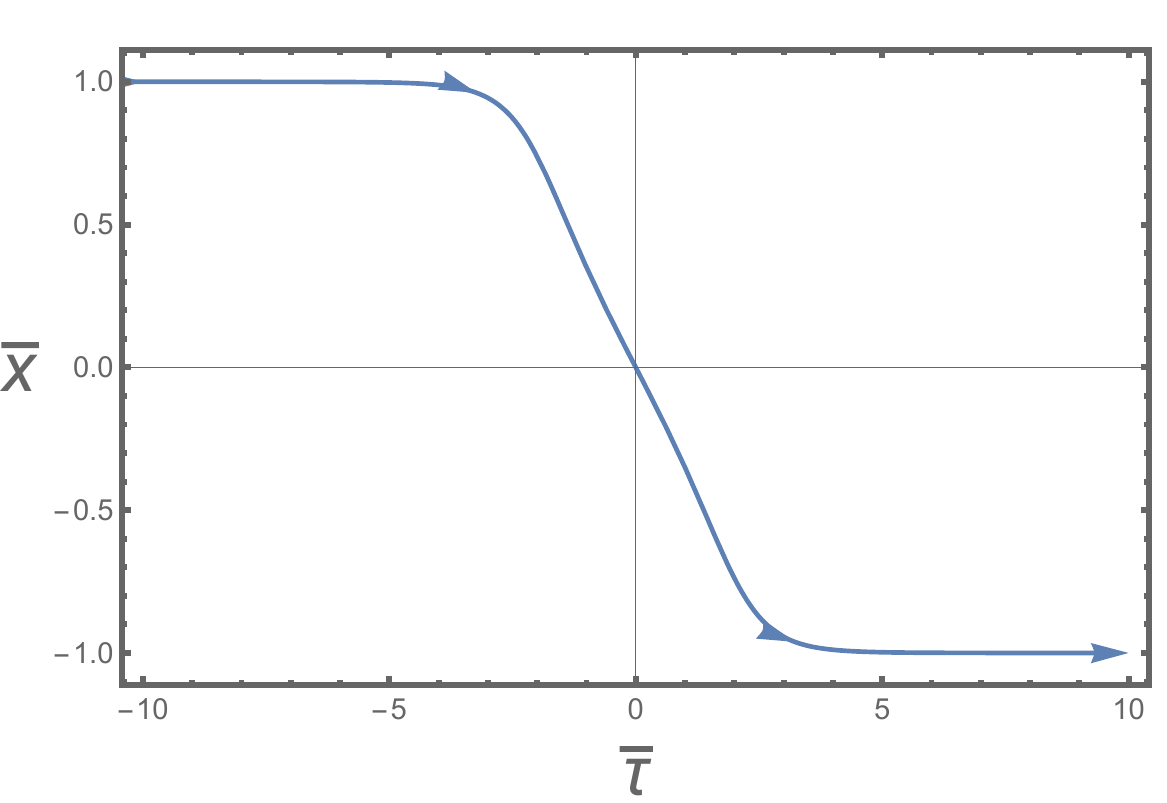}\hspace{1cm}
\includegraphics[scale=0.35]{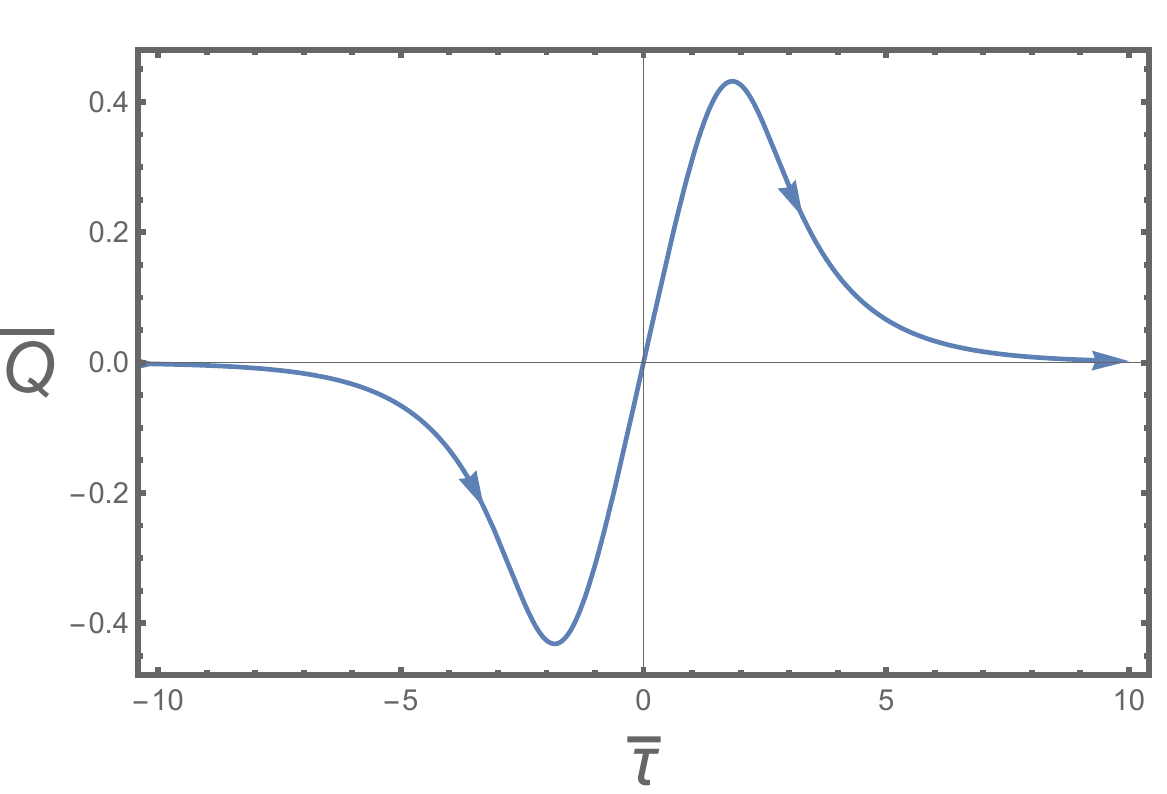}\vspace{1cm}
\caption{ The left and right panels depict the time evolution of the variables $\Bar{x}$ and $\Bar{Q}$ with respect to $\Bar{\tau}$ for the trajectory passing through the origin. $\alpha=2/3$ is considered.}\label{fig:2D_sys_plots_Ex1}
\end{figure}

For the special case when the bounce occurs in a spatially flat and vacuum scenario, $\Bar{\Omega}=0$ and $\Bar{K}=0$, the dynamical system reduces to 
\begin{subequations}
     \begin{eqnarray}
        \frac{d\Bar{x}}{d\Bar{\tau}}&=&\frac{1}{6}  \Big[2\left(-1+\Bar{x}\Bar{Q}+2\Bar{Q}^{2}-2\Bar{x}^{2}\Bar{Q}^{2}-\Bar{x}^{3}\Bar{Q}+\Bar{x}^{4}\right)-\frac{2(2\alpha-3)}{\alpha}\Bar{x}\Bar{Q}(1-\Bar{x}^{2})+\frac{4\Bar{x}\Bar{Q}^{3}}{\alpha^{2}}\left(2\alpha^{2}-3\alpha+1\right)\Big]\,, \notag
        \\
        && 
        \\
        \frac{d\Bar{Q}}{d\Bar{\tau}}&=&\frac{1}{6} \Big[2\Bar{x}\Bar{Q} -4\Bar{Q}^{2} +\Big(2-\frac{2(2\alpha-3)\Bar{Q}^{2}}{\alpha}\Big)(1-\Bar{x}^{2}) +\frac{4\left(2\alpha^{2} -3\alpha+1\right)\Bar{Q}^{4}}{\alpha^{2}}-2\Bar{x}^{2}\Bar{Q}^{2}+2\Bar{x}^{3}\Bar{Q} -4\Bar{x}\Bar{Q}^{3}\Big]\,.\notag
        \\
        &&
    \end{eqnarray}
\end{subequations}
In this case, the phase space is 2-dimensional. The fixed points are presented in Table \ref{tab:crit_model_I}. 
\begin{table}[H]
    \centering
    \begin{tabular}{|c|c|c|c|c|}
    \hline
 Fixed points & Coordinates $(\Bar{x},\Bar{Q})$ & Jacobian eigenvalues & Stability & Cosmology \\ 
 \hline
  $A_{1\pm}$ & $(\pm1,0)$ & $\left\{\pm \frac{4}{3},\pm \frac{2}{3}\right\}$ & \begin{tabular}{@{}c@{}} $(-)$ attractor \\ $(+)$ repeller \end{tabular} & Indeterminate \\ 
  \hline
  $A_{2\pm}$ & $(\pm 0.53,\mp 1.20)$ & $\big\{\mp1.71, \pm 1.12\big\}$ & Saddle & $a(t)\sim|t|^{\frac{2}{3}}$ \\ 
  \hline
  $A_{3\pm}$ & $(\pm 0.923,\pm 0.545)$ & $\{\mp 0.78,\mp 0.54\}$ & \begin{tabular}{@{}c@{}} $(+)$ attractor \\ $(-)$ repeller \end{tabular} & $a(t)\sim|t|^{\frac{2}{3}}$ \\ 
  \hline
  \end{tabular}
    \caption{Fixed points, Jacobian eigenvalues, stability and the respective cosmologies corresponding to the ansatz \eqref{eq:model-I} with $\alpha=\frac{2}{3}$ for the spatially flat vacuum case.}
    \label{tab:crit_model_I}
\end{table}
The phase portrait is presented in Figure \ref{fig:model_I}, where we also highlight the physically viable region according to the condition \eqref{v_cond}. One can note that, in general, all the non-singular bouncing trajectories do go out of the physically viable region for some time during their course of evolution, except for the single trajectory passing through the origin. All other bouncing trajectories encounter either ghost or tachyonic instability, at least in the vicinity of the bounce. Therefore, one can conclude that even though globally stable non-singular bouncing solutions of the form \eqref{eq:model-I} are in principle possible in $f(R)$ gravity, they are very fine-tuned.
The time evolution of the dynamical variables for the trajectory passing through the origin is shown in Figure \ref{fig:2D_sys_plots_Ex1}.

The $f(R)$ phase space corresponding to the ansatz \eqref{eq:model-I} has two invariant submanifolds corresponding to $q = \frac{1-\alpha}{\alpha},\,\frac{2-\alpha}{\alpha}$. This can be seen clearly by inserting the expression of $j=j(q)$ from Eq.\eqref{eq:j-I} into the first equation of \eqref{CP_rel}
\begin{equation}
   \frac{dq}{d\tau} = 2\left(q-\frac{1-\alpha}{\alpha}\right)\left(q-\frac{2-\alpha}{\alpha}\right)\,.
\end{equation}
They represent 3-dimensional surfaces in the 4-dimensional phase space spanned by the variables $\{\bar{x},\Bar{\Omega},\Bar{Q},\Bar{K}\}$, whose equation can be found using Eq.\eqref{eq:dec_non_vacuum}. The intersection of the submanifold $q=\frac{1-\alpha}{\alpha}$ with the spatially flat vacuum invariant submanifold for $\alpha=\frac{2}{3}$ is shown in Figure \ref{fig:model_I} as the green curve. As can be seen from the figure, the bouncing trajectories corresponding to Eq.\eqref{eq:model-I} are contained within the region bounded by this submanifold. Moreover, the submanifold is asymptotic to all the bouncing trajectories, which is consistent with the fact that the past and future asymptotic state of the cosmology \eqref{eq:model-I} have a power law evolution $a(t)\sim|t|^\alpha$ (recall that we have taken the same numerical value for $\beta$ and $\alpha$ for simplicity). The trajectories that lie outside the region bounded by the invariant submanifold $q=\frac{1-\alpha}{\alpha}$ are other possible (non-bouncing) cosmological solutions of the $f(R)$ theory that could have been reconstructed starting from the ansatz \eqref{eq:model-I} (more discussion on this in section \ref{sec:discussion}). The other invariant submanifold $q=\frac{2-\alpha}{\alpha}$ lies outside the compact region and is not of any physical interest.   

\subsection{Example-II}
The second example that we consider has a scale factor that evolves according to
\begin{equation}\label{eq:model-II}
    a(t) = a_0\cosh{\lambda t}\,,    
\end{equation}
where again $-\infty<t<\infty$ and $\lambda$ is a positive constant with $[\lambda]=[t]^{-1}$. In asymptotic past and asymptotic future, the behaviour of the scale factor is $e^{-\lambda t}$ and $e^{\lambda t}$ respectively. Therefore, this ansatz represents a cosmic evolution that connects a contracting De-Sitter (deflationary) phase to an expanding de-Sitter (inflationary) phase through a non-singular bounce. This ansatz was introduced in Ref.\cite{Bouhmadi-Lopez:2012piq} as a model of inflation preceded by a bounce. The authors in Ref.\cite{Bouhmadi-Lopez:2012piq} try to reconstruct the $f(R)$, which comes out in a complicated form. Our approach presents a way around the cumbersome exercise of reconstruction.

Hubble and deceleration parameters for the above-mentioned scale factor are \begin{subequations}
    \begin{eqnarray}
    H&=& \lambda \tanh{{\lambda t}}\,,\label{eq:Hm2}\\
    q&=& -1-\csch^{2}{\lambda t}\,.\label{eq:qm2}
\end{eqnarray}  
\end{subequations}
Eq.\eqref{eq:Hm2} can be inverted to get
\begin{equation}
    \csch^{2}{\lambda t} = -1-q \,.
\end{equation}
The jerk parameter can be calculated from Eq.\eqref{eq:jerk}. Using the above expression for $\csch^{2}{\lambda t}$, we obtain
\begin{equation}\label{eq:j-II}
    j = - q \,.
\end{equation}
Finally, we obtain
\begin{equation}
    \Bar{Q}^{3}(j-q-2)=-2\Bar{Q}^{3}(q+1)=2\Bar{Q}(1-\Bar{\Omega}-\Bar{x}^{2}-\Bar{K})-4\Bar{Q}^{3}  \,,  
\end{equation}
which gives us the dynamical system
\begin{subequations}
    \begin{eqnarray}
        \frac{d\Bar{x}}{d\Bar{\tau}}&=&\frac{1}{6}  \Big[2\left(-1+2\Bar{K}+\Bar{x}\Bar{Q}-2\Bar{x}^{2}\Bar{K}+2\Bar{Q}^{2}-2\Bar{x}^{2}\Bar{Q}^{2}-\Bar{x}^{3}\Bar{Q}+\Bar{x}^{4}\right)-\Bar{\Omega}(1+3\omega)+3\Bar{x}\Bar{\Omega}(1+\omega)(\Bar{Q}+\Bar{x}) \nonumber \\
        &&-2\Bar{x}\Bar{Q}(1-\Bar{\Omega}-\Bar{x}^{2}-\Bar{K})+4\Bar{x}\Bar{Q}^{3}+2\Bar{x}\Bar{K}\Bar{Q}\Big]\,,
        \\
        \frac{d\Bar{\Omega}}{d\Bar{\tau}}&=&\frac{\Bar{\Omega}}{3}\Big[3(\Bar{x}\Bar{\Omega}+\Bar{Q}\Bar{\Omega}-\Bar{Q})(1+\omega)-2\Bar{Q}(1-\Bar{\Omega}-\Bar{x}^{2}-\Bar{K}) +4\Bar{Q}^{3}+2\Bar{K}\Bar{Q}-4\Bar{x}\Bar{Q}^{2}-4\Bar{x}\Bar{K}-2\Bar{x}^{2}\Bar{Q} \nonumber \\
        && +2\Bar{x}^{3}\Big]\,, 
        \\   
        \frac{d\Bar{Q}}{d\Bar{\tau}}&=&\frac{1}{6} \Big[3\Bar{Q}\Bar{\Omega}(1+\omega)(\Bar{Q}+\Bar{x})-2\Bar{K}-4\Bar{Q}^{2}+2\Bar{x}\Bar{Q}-2\Bar{Q}^{2}(1-\Bar{\Omega}-\Bar{x}^{2}-\Bar{K})+4\Bar{Q}^{4}+2\Bar{K}\Bar{Q}^{2} \nonumber \\ &&+2\left(1-\Bar{\Omega} -\Bar{x}^{2}\right)-4\Bar{x}\Bar{Q}\Bar{K} -2\Bar{x}^{2}\Bar{Q}^{2} +2\Bar{x}^{3}\Bar{Q}-4\Bar{x}\Bar{Q}^{3}\Big]\,, 
        \\
        \frac{d\Bar{K}}{d\Bar{\tau}}&=&\frac{\Bar{K}}{6} \Big[4\Bar{x}-4\Bar{Q}-8\Bar{x}\Bar{K}-8\Bar{x}\Bar{Q}^{2}-4\Bar{x}^{2}\Bar{Q}+4\Bar{x}^{3}+6\Bar{\Omega}(1+\omega)(\Bar{Q}+\Bar{x})-4\Bar{Q}(1-\Bar{\Omega}-\Bar{x}^{2}-\Bar{K}) \nonumber \\ &&+8\Bar{Q}^{3}+4\Bar{K}\Bar{Q}\Big]\,.
    \end{eqnarray}    
\end{subequations}
The fixed points of this dynamical system are listed in Table \ref{tab:generic_crit_model_IIa} and their stability
and respective cosmologies are listed in Table \ref{tab:generic_crit_model_IIaa}.
\begin{table}[H]
\resizebox{\textwidth}{!}{
    \centering
    \begin{tabular}{|c|c|c|c|}
    \hline
 \begin{tabular}{@{}c@{}} Fixed \\ points \end{tabular} & Coordinates $(\Bar{x},\Bar{\Omega},\Bar{Q},\Bar{K})$ & Existence & Jacobian eigenvalues  \\ 
\hline
  $B_{1\pm}$ & $(\pm1,0,0,0)$ & Always & $\left\{\pm \frac{4}{3},\pm \frac{4}{3},\pm \frac{2}{3},\pm \frac{2}{3}\right\}$  \\ 
\hline
  $B_{2\pm}$ & $\left(0,0,\pm\frac{1}{\sqrt{2}},0\right)$ & Always & $\left\{\mp \frac{1}{\sqrt{2}},\mp \frac{\sqrt{2}}{3},\mp \frac{\sqrt{2}}{3},\pm \frac{1}{3 \sqrt{2}}\right\}$  \\ 
\hline
  $B_{3\pm}$ & $\left(\pm\frac{1}{3},0,\pm\frac{2}{3},0\right)$ & Always & $\left\{\mp \frac{8}{9},\mp \frac{4}{9},\mp \frac{4}{9},\mp \frac{2}{9}\right\}$  \\ 
\hline
  $B_{4+}$ & $\left(\frac{\Bar{Q}+ \sqrt{3+\Bar{Q}^{2}}}{3},0,\Bar{Q},\frac{6-11\Bar{Q}^{2}- 2\Bar{Q}\sqrt{3+\Bar{Q}^{2}}}{9}\right)$ & $-0.939\leq \Bar{Q} \leq 0.59$ & $\left\{0,\frac{2 \Bar{Q}}{3},\frac{2\left(\Bar{Q}- 2 \sqrt{\Bar{Q}^2+3}\right)}{9},\frac{\left(- 2 \sqrt{\Bar{Q}^2+3}-5 \Bar{Q}\right)}{9} \right\}$  \\ 
\hline
  $B_{4-}$ & $\left(\frac{\Bar{Q}- \sqrt{3+\Bar{Q}^{2}}}{3},0,\Bar{Q},\frac{6-11\Bar{Q}^{2}+ 2\Bar{Q}\sqrt{3+\Bar{Q}^{2}}}{9}\right)$ & $-0.59\leq \Bar{Q}\leq 0.939$ & $\left\{0,\frac{2 \Bar{Q}}{3},\frac{2\left(\Bar{Q}+ 2 \sqrt{\Bar{Q}^2+3}\right)}{9},\frac{\left(2 \sqrt{\Bar{Q}^2+3}-5 \Bar{Q}\right)}{9} \right\}$  \\ 
\hline
  $B_{5}$ & $\left(-\frac{\Bar{Q}}{2},\frac{4-7\Bar{Q}^{2}}{10},\Bar{Q},\frac{12-11\Bar{Q}^{2}}{20}\right)$ & $-\frac{2}{\sqrt{7}}\leq \Bar{Q} \leq \frac{2}{\sqrt{7}}$ & $\left\{0,\frac{2 \Bar{Q}}{3},\frac{2 \Bar{Q}}{3}-\frac{1}{3} \sqrt{\frac{19 \Bar{Q}^2}{10}+\frac{6}{5}},\frac{\left(\sqrt{190 \Bar{Q}^2+120}+20 \Bar{Q}\right)}{30}\right\}$  \\ 
\hline
  \end{tabular}}
    \caption{Fixed points, their existence conditions and the Jacobian eigenvalues for the cosmic evolution corresponding to the ansatz \eqref{eq:model-II} with $\omega=0$.}
    \label{tab:generic_crit_model_IIa}
\end{table}
\begin{table}[H]
    \centering
    \begin{tabular}{|c|c|c|c|}\hline
 Fixed points & Coordinates $(\Bar{x},\Bar{\Omega},\Bar{Q},\Bar{K})$ & Stability & Cosmology \\ \hline
  $B_{1\pm}$ & $(\pm1,0,0,0)$ & \begin{tabular}{@{}c@{}} $(-)$ attractor\\ $(+)$ repeller \end{tabular} & Indeterminate \\ 
  \hline
  $B_{2\pm}$ & $\left(0,0,\pm\frac{1}{\sqrt{2}},0\right)$ & Saddle & \begin{tabular}{@{}c@{}} $a(t)=a_{0}e^{ct}$ \\ (De-sitter) \end{tabular} \\ 
  \hline
  $B_{3\pm}$ & $\left(\pm\frac{1}{3},0,\pm\frac{2}{3},0\right)$ & \begin{tabular}{@{}c@{}} $(+)$ attractor\\ $(-)$ repeller \end{tabular} & \begin{tabular}{@{}c@{}} $a(t)=a_{0}e^{ct}$ \\ (De-sitter) \end{tabular} \\ 
  \hline 
  $B_{4+}$ & $\left(\frac{\Bar{Q} + \sqrt{3+\Bar{Q}^{2}}}{3},0,\Bar{Q},\frac{6-11\Bar{Q}^{2} - 2\Bar{Q}\sqrt{3+\Bar{Q}^{2}}}{9}\right)$ & \begin{tabular}{@{}c@{}} Stable for $-\frac{2}{\sqrt{7}}<\Bar{Q}<0$ \\ saddle otherwise \end{tabular} & $a(t)=a_{0}+Ct$  \\ 
  \hline 
  $B_{4-}$ & $\left(\frac{\Bar{Q} - \sqrt{3+\Bar{Q}^{2}}}{3},0,\Bar{Q},\frac{6-11\Bar{Q}^{2} + 2\Bar{Q}\sqrt{3+\Bar{Q}^{2}}}{9}\right)$ & \begin{tabular}{@{}c@{}} Unstable for $0<\Bar{Q}<\frac{2}{\sqrt{7}}$ \\ saddle otherwise \end{tabular} & $a(t)=a_{0}+Ct$  \\ 
  \hline
  $B_{5}$ & $\left(-\frac{\Bar{Q}}{2},\frac{4-7\Bar{Q}^{2}}{10},\Bar{Q},\frac{12-11\Bar{Q}^{2}}{20}\right)$ & Saddle & $a(t)=a_{0}+Ct$ \\ 
  \hline
  \end{tabular}
    \caption{Fixed points, stability and the respective cosmologies for the cosmic evolution corresponding to the ansatz \eqref{eq:model-II} with $\omega=0$. For those fixed points which have $\frac{\dot{H}}{H^2}\rightarrow\frac{0}{0}$ on approach to the fixed point, cosmology is indeterminate.}
    \label{tab:generic_crit_model_IIaa}
\end{table}
For the points $B_{1\pm}$, we cannot use the same argument that we used for model-I \eqref{eq:model-I}, as for the model \eqref{eq:model-II}, both $a(t)$ and $\dot{a}(t)$ diverges at the same rate as $|t|\rightarrow\infty$. So, unfortunately, we are unable to determine their cosmology unambiguously. We recognize this as a limitation of our method.

\begin{figure}[H]
	\begin{center}
		\includegraphics[width=7cm]{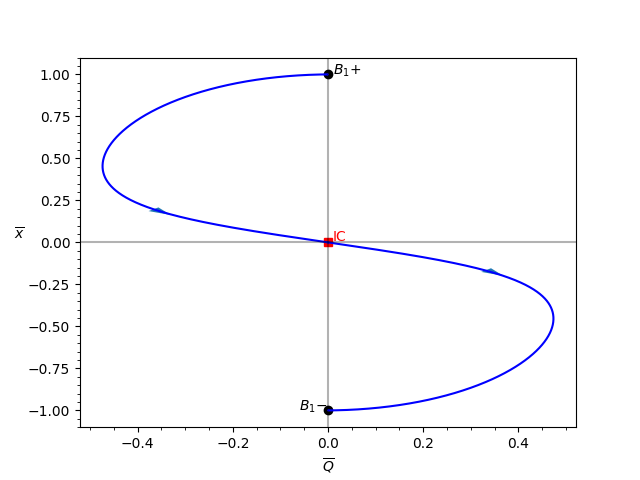}
		\includegraphics[width=7cm]{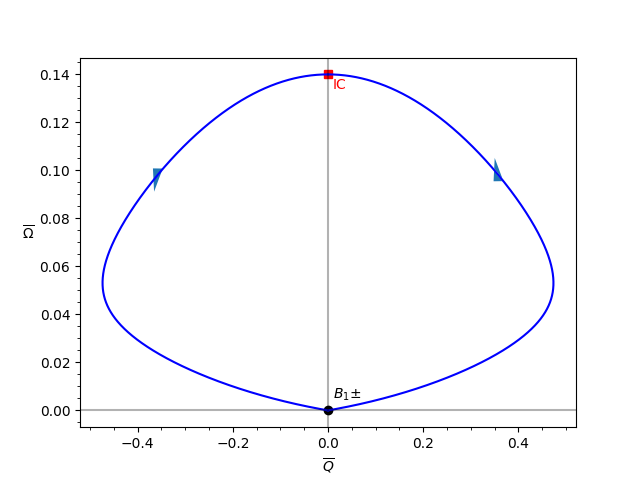}
		\includegraphics[width=7cm]{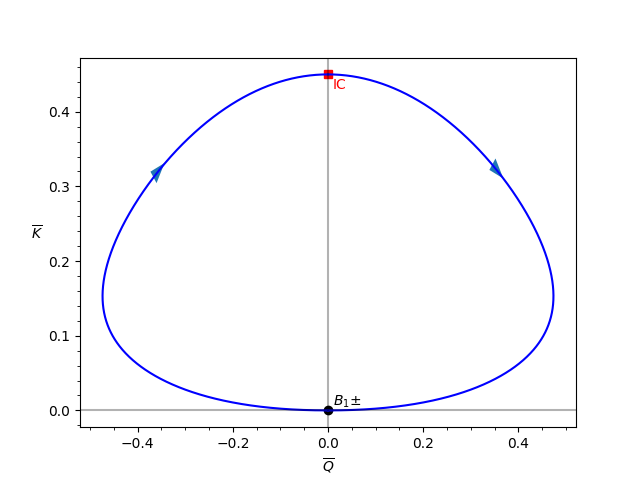}
            \includegraphics[width=7cm]{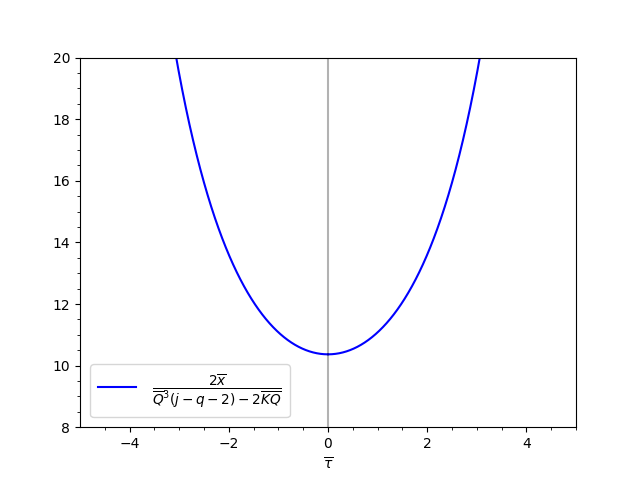}
	\end{center}
	\caption{Parametric plots $\bar{x}(\bar{\tau})$ v/s $\bar{Q}(\bar{\tau})$ (upper left), $\bar{\Omega}(\bar{\tau})$ v/s $\bar{Q}(\bar\tau)$ (upper right) and $\bar{K}(\bar\tau)$ v/s $\bar{Q}(\bar\tau)$ (lower left) for a trajectory passing through the point $(\bar{x},\bar{\Omega},\bar{Q},\bar{K})=(0,0.14,0,0.45)$ for the ansatz \eqref{eq:model-II} with $\omega=0$. The lower right panel shows the quantity $\frac{2\Bar{x}}{\Bar{Q}^{3}(j-q-2)-2\Bar{K}\Bar{Q}}$ remains positive throughout the entire evolution history of the bouncing cosmology, signaling a stable bouncing solution.}
\label{fig:model_II_a}
\end{figure}

Figures \ref{fig:model_II_a} show different projections of two-phase trajectories that correspond to a non-singular bounce. The last panel in the figure proves that throughout its entire evolution history, bouncing cosmology encounters neither ghost nor tachyonic instability.

\begin{figure}[H]
    \centering
    \includegraphics[width=6cm]{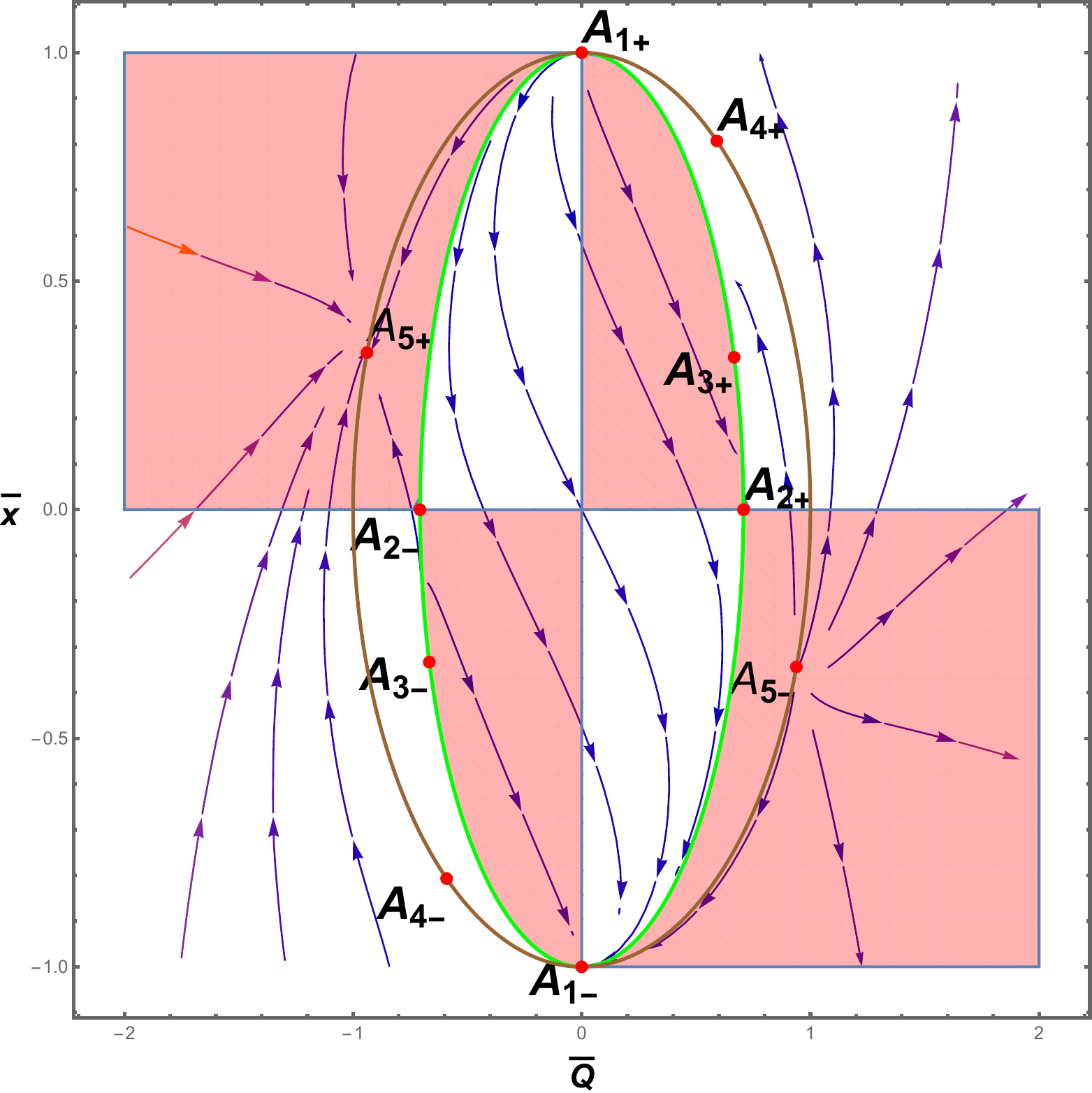}
    \caption{Phase portrait for the spatially flat vacuum case for the ansatz \eqref{eq:model-II}. The shaded region is the physically viable region where the $f(R)$ theory is free from ghost and tachyonic instability ($F>0,\,F'>0$).  The green ellipse corresponds to the invariant submanifold $q=-1$. The brown ellipse corresponds to the invariant submanifold $q=0$.}
    \label{fig:model_II}
\end{figure}

\begin{figure}[H]
\centering
\includegraphics[scale=0.35]{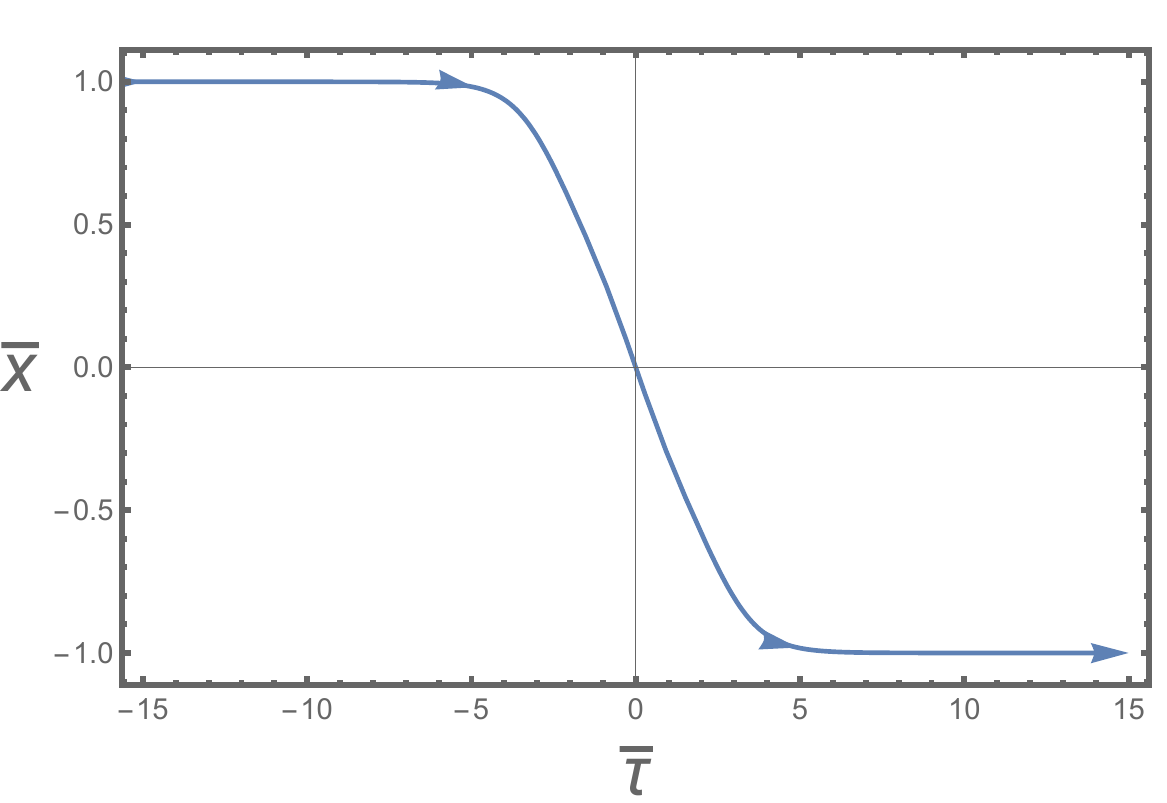}\hspace{1cm}
\includegraphics[scale=0.35]{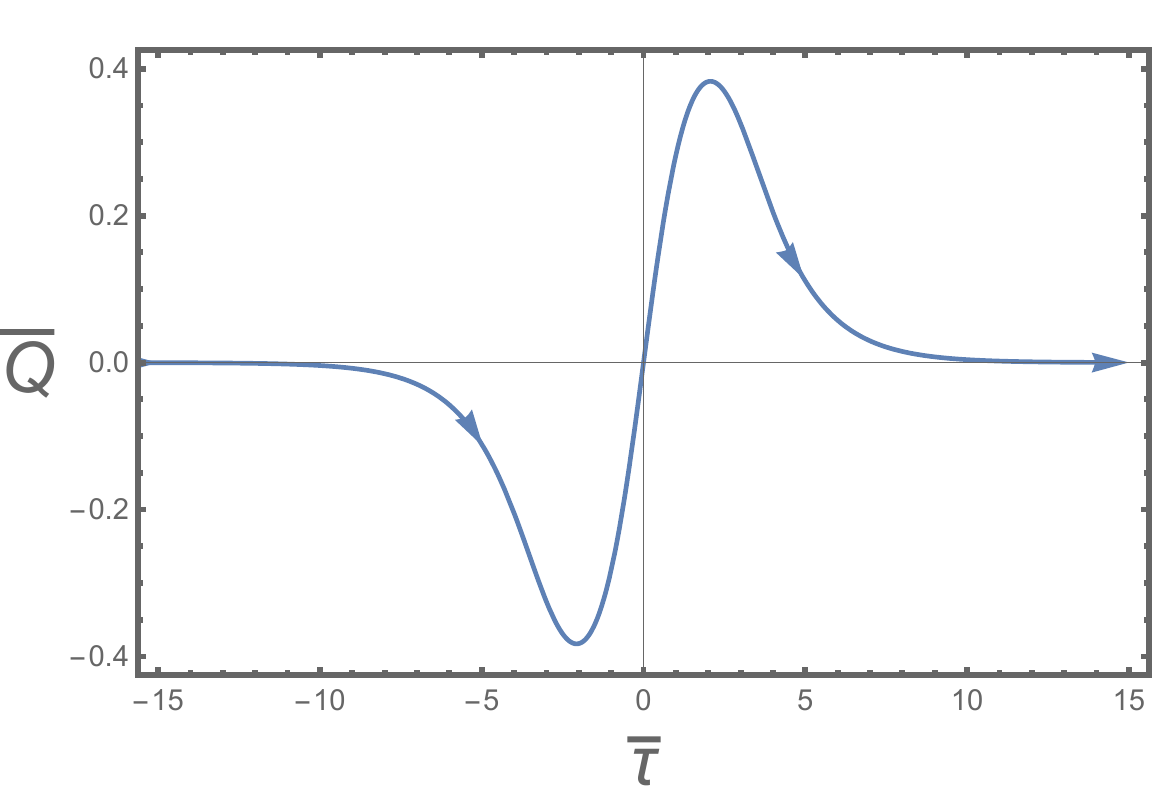}\vspace{1cm}
\caption{The left and right panels, respectively, depict the time evolution of the variables $\Bar{x}$ and $\Bar{Q}$ with respect to $\Bar{\tau}$ for the trajectory passing through the origin.}
\label{fig:2D_sys_plots_Ex2}
\end{figure}

For the spatially flat vacuum case, the dynamical system reduces to
\begin{subequations}
    \begin{eqnarray}
        \frac{d\Bar{x}}{d\Bar{\tau}}&=&\frac{1}{6}  \Big[2\left(-1+\Bar{x}\Bar{Q}+2\Bar{Q}^{2}-2\Bar{x}^{2}\Bar{Q}^{2}-\Bar{x}^{3}\Bar{Q}+\Bar{x}^{4}\right)-2\Bar{x}\Bar{Q}(1-\Bar{x}^{2})+4\Bar{x}\Bar{Q}^{3}\Big]\,,
        \\
        \frac{d\Bar{Q}}{d\Bar{\tau}}&=&\frac{1}{6} \Big[-4\Bar{Q}^{2}+2\Bar{x}\Bar{Q}+\left(2-2\Bar{Q}^{2}\right)(1-\Bar{x}^{2})+4\Bar{Q}^{4}-2\Bar{x}^{2}\Bar{Q}^{2} +2\Bar{x}^{3}\Bar{Q}-4\Bar{x}\Bar{Q}^{3}\Big]\,. 
        \end{eqnarray}    
\end{subequations}
whose fixed points are listed in Table \ref{tab:crit_model_II}.
\begin{table}[H]
    \centering
    \begin{tabular}{|c|c|c|c|c|}
\hline
 Fixed points & Coordinates $(\Bar{x},\Bar{Q})$ & Jacobian eigenvalues & Stability & Cosmology \\ 
\hline
  $A_{1\pm}$ & $(\pm 1,0)$ &  $\left\{\pm \frac{4}{3},\pm \frac{2}{3}\right\}$ & \begin{tabular}{@{}c@{}}$(-)$ attractor\\ $(+)$ repeller \end{tabular} & Indeterminate \\ 
\hline
  $A_{2\pm}$ & $\left(0,\pm\frac{1}{\sqrt{2}}\right)$ & $\left\{\mp \frac{\sqrt{2}}{3},\pm \frac{1}{3 \sqrt{2}}\right\}$ & Saddle & \begin{tabular}{@{}c@{}} $a(t)=a_{0}e^{ct}$ \\ (De-sitter) \end{tabular} \\ 
\hline
  $A_{3\pm}$ & $\left(\pm\frac{1}{3},\pm\frac{2}{3}\right)$ & $\left\{\mp \frac{4}{9},\mp \frac{2}{9}\right\}$ & \begin{tabular}{@{}c@{}} $(+)$ attractor\\ $(-)$ repeller \end{tabular} & \begin{tabular}{@{}c@{}} $a(t)=a_{0}e^{ct}$ \\ (De-sitter) \end{tabular} \\ 
\hline
  $A_{4\pm}$ & $(\approx \pm 0.81,\approx \pm 0.59)$ & $\{\mp 0.68, \pm 0.39\}$ & Saddle & $a(t)=a_{0}+Ct$ \\ 
\hline
  $A_{5\pm}$ & $(\approx \pm 0.34, \approx \mp 0.94)$ & $\{\mp 1.08,\mp 0.625\}$ & \begin{tabular}{@{}c@{}} $(+)$ attractor\\ $(-)$ repeller \end{tabular} & $a(t)=a_{0}+Ct$ \\ 
\hline
  \end{tabular}
    \caption{Fixed points, Jacobian eigenvalues, stabilities and the respective cosmologies for the ansatz \eqref{eq:model-II} for the spatially flat vacuum case.}
    \label{tab:crit_model_II}
\end{table}
The phase portrait is presented in Figure \ref{fig:model_II}, where the physically viable region is highlighted. Amazingly, one can notice that there are no bouncing trajectories that remain completely within the physically viable region for the entire course of evolution. If one considers a spatially flat evolution of the form \eqref{eq:model-II} in a vacuum scenario, within the framework of $f(R)$ gravity, it is bound to encounter either ghost or tachyonic instability at least in the vicinity of the bounce. This makes the model considered in Ref.\cite{Bouhmadi-Lopez:2012piq} not physically viable.
The time evolution of the dynamical variables for the trajectory passing through the origin is shown in Figure \ref{fig:2D_sys_plots_Ex2}.

The $f(R)$ phase space corresponding to the ansatz \eqref{eq:model-II} has two invariant submanifolds corresponding to $q = 0,\,-1$. This can be seen clearly by inserting the expression of $j=j(q)$ from Eq.\eqref{eq:j-II} into the first equation of \eqref{CP_rel}
\begin{equation}
   \frac{dq}{d\tau} = 2q(q+1)\,.
\end{equation}
The equations of the corresponding 3-dimensional surfaces can be found using Eq.\eqref{eq:dec_non_vacuum}. The intersections of these invariant submanifolds with the spatially flat vacuum invariant submanifold are shown in Figure \ref{fig:model_II} as the green and brown curves. As can be seen from the figure, the bouncing trajectories corresponding to Eq.\eqref{eq:model-II} are contained within the region bounded by the submanifold $q=-1$. This submanifold is asymptotic to all the bouncing trajectories, which is consistent with the fact that the past and future asymptotic states of the cosmology \eqref{eq:model-II} are De-Sitter. The trajectories that lie outside the region bounded by the invariant submanifold $q=-1$ are other possible (non-bouncing) cosmological solutions of the $f(R)$ theory that could have been reconstructed starting from the ansatz \eqref{eq:model-II} (more discussion on this in section \ref{sec:discussion}).   

\subsection{Example III}
The third and the last example that we consider is a non-singular cyclic cosmology
\begin{equation}\label{eq:model-III}
a(t)=\frac{1}{2\Lambda}\left[2-\cos(2\zeta t)\right] \,,  
\end{equation}
where $\Lambda$ is a dimensionless constant and $\zeta$ is a constant with dimension $[\zeta]=[t]^{-1}$. It has been argued that a cyclic cosmological model is compatible with current observations if the cosmological constant happens to be decaying \cite{Ellis:2015bag}. The simple example we use here is taken from Ref.\cite{Pavlovic:2020sei}, where the authors investigated whether such a cyclic cosmology can be supported by a third-order polynomial form of $f(R)$. The authors in Ref.\cite{Pavlovic:2020sei} did not investigate the stability of such a model. As we will see below, a simple cyclic ansatz like \eqref{eq:model-III} is totally unstable in $f(R)$ gravity, irrespective of its form. 

The deceleration and the jerk parameter is calculated from Eq.\eqref{eq:decel} and Eq.\eqref{eq:jerk} respectively
\begin{subequations}
\begin{eqnarray}
q &=& \frac{\cos^{2}(2\zeta t)-2\cos(2\zeta t)}{\sin^{2}(2\zeta t)}\,,
\\
j &=& -[\cot (2\zeta t)-2 \csc (2\zeta t)]^2=\frac{\cos^{2}(2\zeta t)-4}{1-\cos^{2}(2\zeta t)}-2q \,.
\end{eqnarray}
\end{subequations}
From the expression of the deceleration parameter above one can solve
\begin{equation}
\cos(2\zeta t) = \frac{1-\sqrt{1+q+q^{2}}}{1+q}\,,    
\end{equation}
using which one arrives at the expression
\begin{equation}
j = -2 \left(\sqrt{q^2+q+1}+1\right)-q \,.
\end{equation}
Finally, one obtains
\begin{equation}
    \Bar{Q}^{3}(j-q-2)=-2 \Bar{Q} \left( \sqrt{{\left(\bar{K}+\bar{x}^2+\bar{\Omega} -1\right) \left(\bar{K}+3 \bar{Q}^2+\bar{x}^2+\bar{\Omega} -1\right)}+3\bar{Q}^4}+\bar{K}+3 \bar{Q}^2+\bar{x}^2+\bar{\Omega} -1\right)\,.    
\end{equation}
We are now in a position to write the dynamical system
\begin{subequations}
    \begin{eqnarray}
        \frac{d\Bar{x}}{d\Bar{\tau}}&=&\frac{1}{6}  \Big[2\left(-1+2\Bar{K}+\Bar{x}\Bar{Q}-2\Bar{x}^{2}\Bar{K}+2\Bar{Q}^{2}-2\Bar{x}^{2}\Bar{Q}^{2}-\Bar{x}^{3}\Bar{Q}+\Bar{x}^{4}\right)-\Bar{\Omega}(1+3\omega)+3\Bar{x}\Bar{\Omega}(1+\omega)(\Bar{Q}+\Bar{x})\nonumber \\
        &&+2\Bar{x}\Bar{K}\Bar{Q} +2 \Bar{x} \Bar{Q} \left( \sqrt{{\left(\bar{K}+\bar{x}^2+\bar{\Omega} -1\right) \left(\bar{K}+3 \bar{Q}^2+\bar{x}^2+\bar{\Omega} -1\right)}+3\bar{Q}^4}+\bar{K}+3 \bar{Q}^2+\bar{x}^2+\bar{\Omega} -1\right)\Big],
        \nonumber \\ 
        \\
        \frac{d\Bar{\Omega}}{d\Bar{\tau}}&=&\frac{\Bar{\Omega}}{3}\Big[3(\Bar{x}\Bar{\Omega}+\Bar{Q}\Bar{\Omega}-\Bar{Q})(1+\omega)+2\Bar{K}\Bar{Q}-4\Bar{x}\Bar{Q}^{2}-4\Bar{x}\Bar{K}-2\Bar{x}^{2}\Bar{Q} +2\Bar{x}^{3} \nonumber \\
        && +2 \Bar{Q} \left( \sqrt{{\left(\bar{K}+\bar{x}^2+\bar{\Omega} -1\right) \left(\bar{K}+3 \bar{Q}^2+\bar{x}^2+\bar{\Omega} -1\right)}+3\bar{Q}^4}+\bar{K}+3 \bar{Q}^2+\bar{x}^2+\bar{\Omega} -1\right)\Big], 
        \\  
        \frac{d\Bar{Q}}{d\Bar{\tau}}&=&\frac{1}{6} \Big[3\Bar{Q}\Bar{\Omega}(1+\omega)(\Bar{Q}+\Bar{x})-2\Bar{K}-4\Bar{Q}^{2}+2\Bar{x}\Bar{Q}+2\Bar{K}\Bar{Q}^{2} +2\left(1-\Bar{\Omega} -\Bar{x}^{2}\right)-4\Bar{x}\Bar{Q}\Bar{K} -2\Bar{x}^{2}\Bar{Q}^{2}+2\Bar{x}^{3}\Bar{Q}  \nonumber\\ 
        &&-4\Bar{x}\Bar{Q}^{3}+2\bar{Q}^{2} \left( \sqrt{{\left(\bar{K}+\bar{x}^2+\bar{\Omega} -1\right) \left(\bar{K}+3 \bar{Q}^2+\bar{x}^2+\bar{\Omega} -1\right)}+3\bar{Q}^4}+\bar{K}+3 \bar{Q}^2+\bar{x}^2+\bar{\Omega} -1\right)\Big], \nonumber \\
        \\
        \frac{d\Bar{K}}{d\Bar{\tau}}&=&\frac{\Bar{K}}{6} \Big[4\Bar{x}-4\Bar{Q}-8\Bar{x}\Bar{K}-8\Bar{x}\Bar{Q}^{2}-4\Bar{x}^{2}\Bar{Q}+4\Bar{x}^{3}+6\Bar{\Omega}(1+\omega)(\Bar{Q}+\Bar{x})+4\Bar{K}\Bar{Q}\nonumber \\
        &&+4 \bar{Q} \left( \sqrt{{\left(\bar{K}+\bar{x}^2+\bar{\Omega} -1\right) \left(\bar{K}+3 \bar{Q}^2+\bar{x}^2+\bar{\Omega} -1\right)}+3\bar{Q}^4}+\bar{K}+3 \bar{Q}^2+\bar{x}^2+\bar{\Omega} -1\right)\Big].
    \end{eqnarray}    
\end{subequations}
An important point is worth mentioning here. The method we devised in this paper depends crucially on the deceleration and the jerk parameter, both of which are periodic for this model. One has to be careful while dealing with periodic functions. The fixed point structure that will be presented below strictly holds within one period $t\in\left[-\frac{\pi}{\zeta},\frac{\pi}{\zeta}\right]$.

The fixed points are listed in Table \ref{tab:generic_crit_model_IIIb}. 
\begin{table}[H]
\resizebox{\textwidth}{!}{
    \centering
    \begin{tabular}{|c|c|c|c|c|}
\hline
 \begin{tabular}{@{}c@{}} Fixed \\ points \end{tabular}  & Coordinates $(\Bar{x},\Bar{\Omega},\Bar{Q},\Bar{K})$ & Jacobian eigenvalues  & Stability & Cosmology \\ 
  \hline
  $B_{1+}$ & $(1,0,0,0)$ &  $\left\{\frac{4}{3},\frac{4}{3},\frac{2}{3},\frac{2}{3}\right\}$ & \begin{tabular}{@{}c@{}} always \\ repeller \end{tabular} & Indeterminate \\ 
  \hline
  $B_{1-}$ & $(-1,0,0,0)$ &  $\left\{- \frac{4}{3},- \frac{4}{3},- \frac{2}{3},- \frac{2}{3}\right\}$ & \begin{tabular}{@{}c@{}} always \\ attractor \end{tabular} & Indeterminate \\ 
  \hline
  $B_{2+}$ & $\left(\frac{1}{\sqrt{3}},0,0,\frac{2}{3}\right)$ & $\{-0.561285,\, 0.0458372,\, 0.116841\, \pm 0.4318 i\}$ & Saddle & Indeterminate \\ 
  \hline
  $B_{2-}$ & $\left(- \frac{1}{\sqrt{3}},0,0,\frac{2}{3}\right)$ & $\{-0.296296,\,-0.133618,\, 0.458974\, \pm 0.50229 i\}$ & Saddle & Indeterminate \\ \hline
  $B_{3}$ & $\left(0,\frac{2}{5},0,\frac{3}{5}\right)$ & $\{0.66674,0.431396,0.158996,-0.017242\}$ & Saddle & Indeterminate \\ 
  \hline
  \end{tabular}}
    \caption{Fixed points, Jacobian eigenvalues and stabilities corresponding to the ansatz \eqref{eq:model-III} with $\omega=0$. Interestingly, all the fixed points have the limiting value $\frac{\dot{H}}{H^2}\rightarrow\frac{0}{0}$ on approach to the fixed point. Therefore, their cosmologies are indeterminate.}
    \label{tab:generic_crit_model_IIIb}
\end{table}
We note that, interestingly, all the fixed points have indeterminate deceleration parameters. We cannot use the argument that we used for model-I \eqref{eq:model-I}, as the cosmology is cyclic. So, unfortunately, we are unable to determine their cosmology unambiguously. We recognize this as a limitation of our method.

\begin{figure}[H]
	\begin{center}
		\includegraphics[width=7cm]{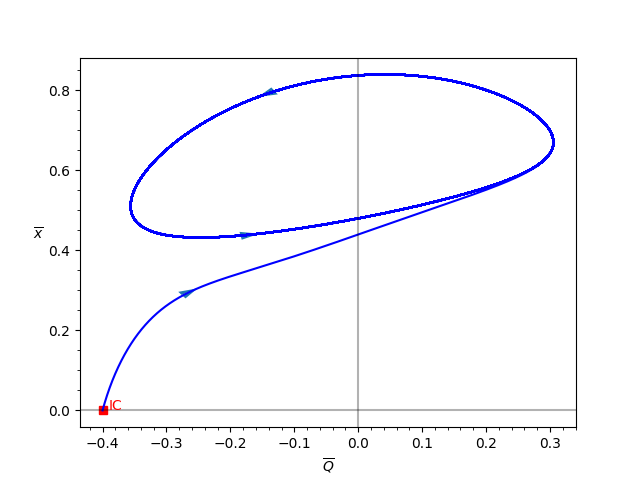}
            \includegraphics[width=7cm]{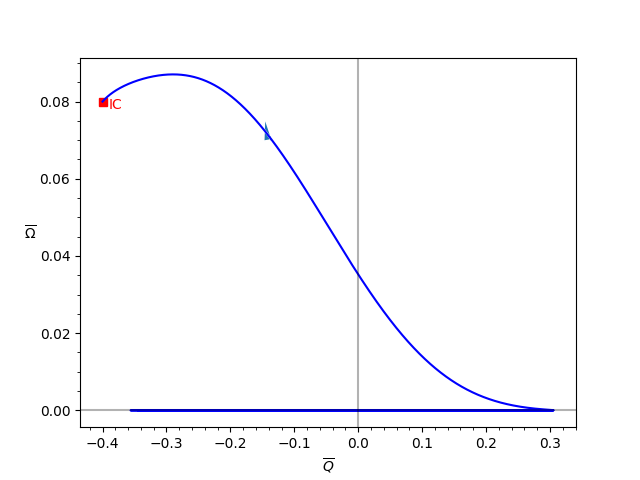}
		\includegraphics[width=7cm]{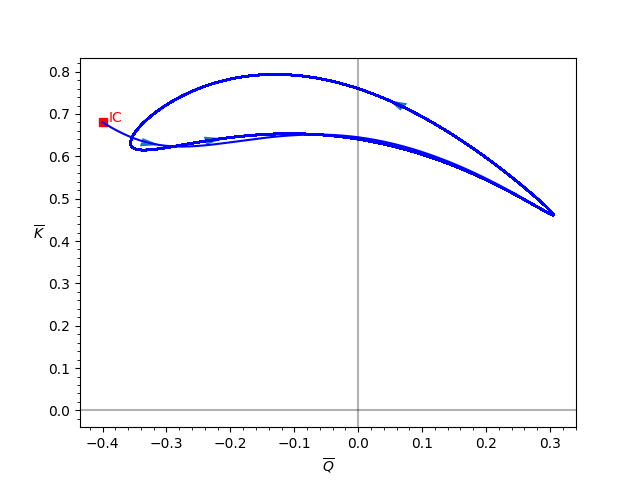}
            \includegraphics[width=7cm]{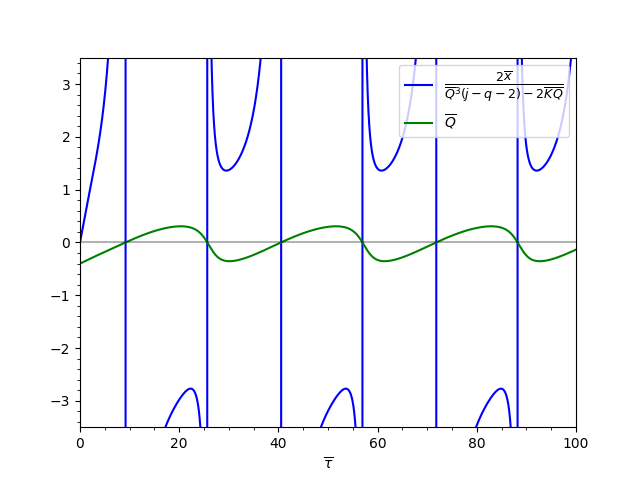}
	\end{center}
	\caption{Parametric plots for the ansatz \eqref{eq:model-III} of $\bar{x}(\bar{\tau})$ v/s $\bar{Q}(\bar{\tau})$ (upper left), $\bar{\Omega}(\bar{\tau})$ v/s $\bar{Q}(\bar\tau)$ (upper right) and $\bar{K}(\bar\tau)$ v/s $\bar{Q}(\bar\tau)$ (lower left) for a trajectory passing through the point $(\bar{x},\bar{\Omega},\bar{Q},\bar{K})=(0,0.08,-0.4,0.68)$. The lower right panel shows the time evolution of $\bar Q$ and the quantity $\frac{2\Bar{x}}{\Bar{Q}^{3}(j-q-2)-2\Bar{K}\Bar{Q}}$ with respect to time variable $\bar{\tau}$.}
\label{model_III_b}
\end{figure}

Parametric plots, however, clearly show the cyclic nature of the cosmology. In Figure \ref{model_III_b}, we show the parametric plots $\bar{x}(\bar{\tau})$ v/s $\bar{Q}(\bar{\tau})$, $\bar{\Omega}(\bar{\tau})$ v/s $\bar{Q}(\bar\tau)$ and $\bar{K}(\bar\tau)$ v/s $\bar{Q}(\bar\tau)$ for a trajectory passing through the point $(\bar{x},\bar{\Omega},\bar{Q},\bar{K})=(0,0.08,-0.4,0.68)$. The last panel of Figure \ref{model_III_b} investigates the physical viability of the cyclic cosmology \eqref{eq:model-III} in $f(R)$ gravity. What the plot reveals is that the expansion phases of the cyclic cosmology are always unstable. This makes a simplistic cyclic model like \eqref{eq:model-III} totally non-viable in $f(R)$ gravity. 

\begin{figure}[H]
    \centering
    \includegraphics[width=6cm]{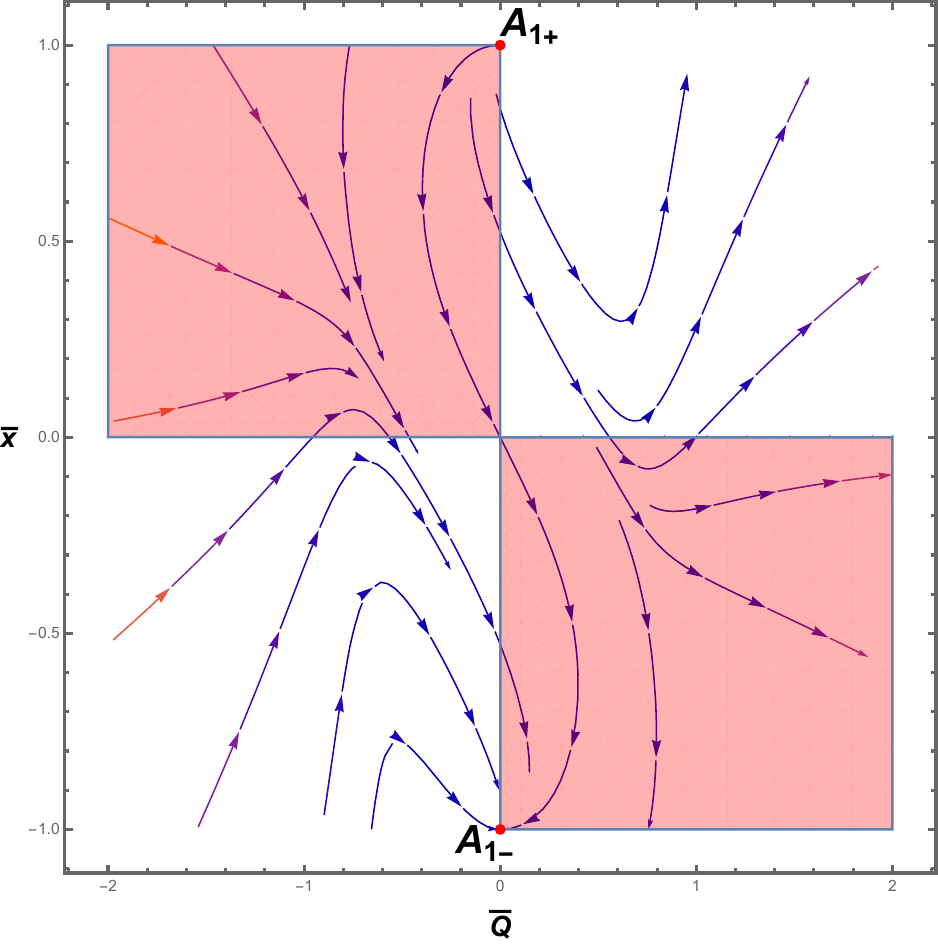}
    \caption{Phase portrait for the spatially flat vacuum case for the ansatz \eqref{eq:model-III}. The shaded region represents absence of ghost and tachyonic instability.}
    \label{fig:model_III}
\end{figure}

\begin{figure}[H]
\centering
\includegraphics[scale=0.35]{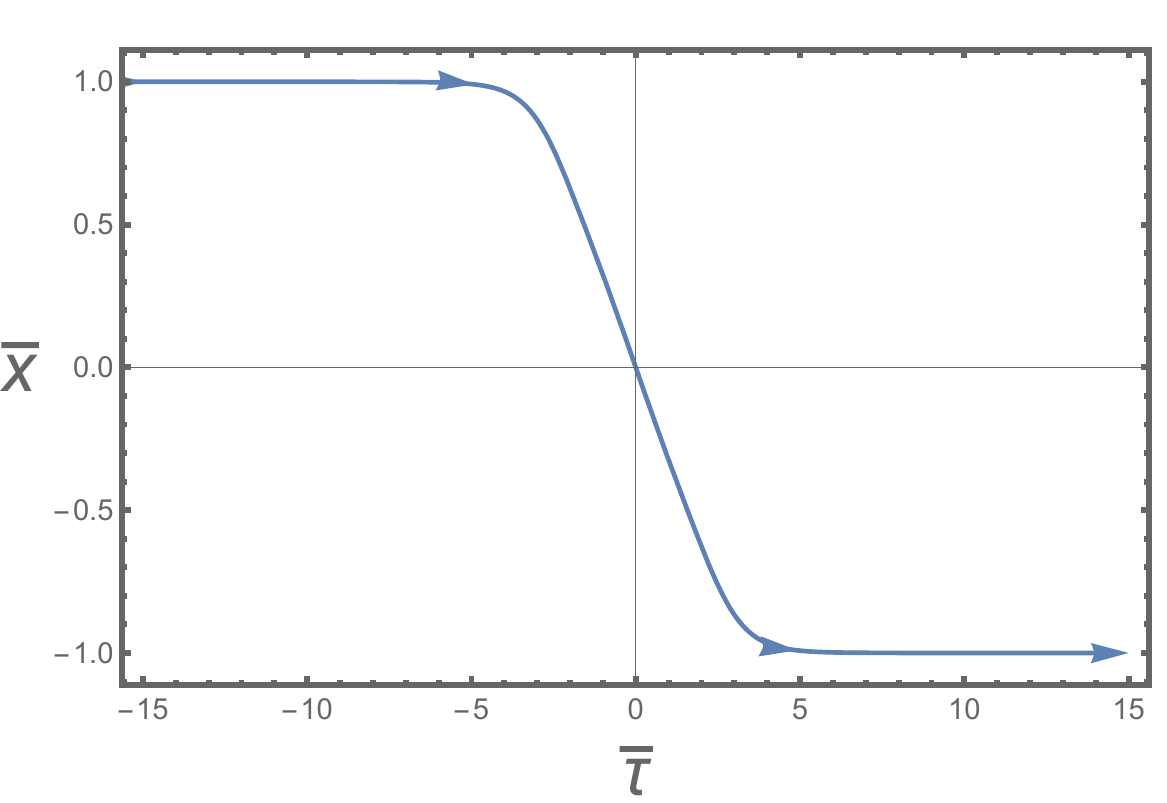}\hspace{1cm}
\includegraphics[scale=0.35]{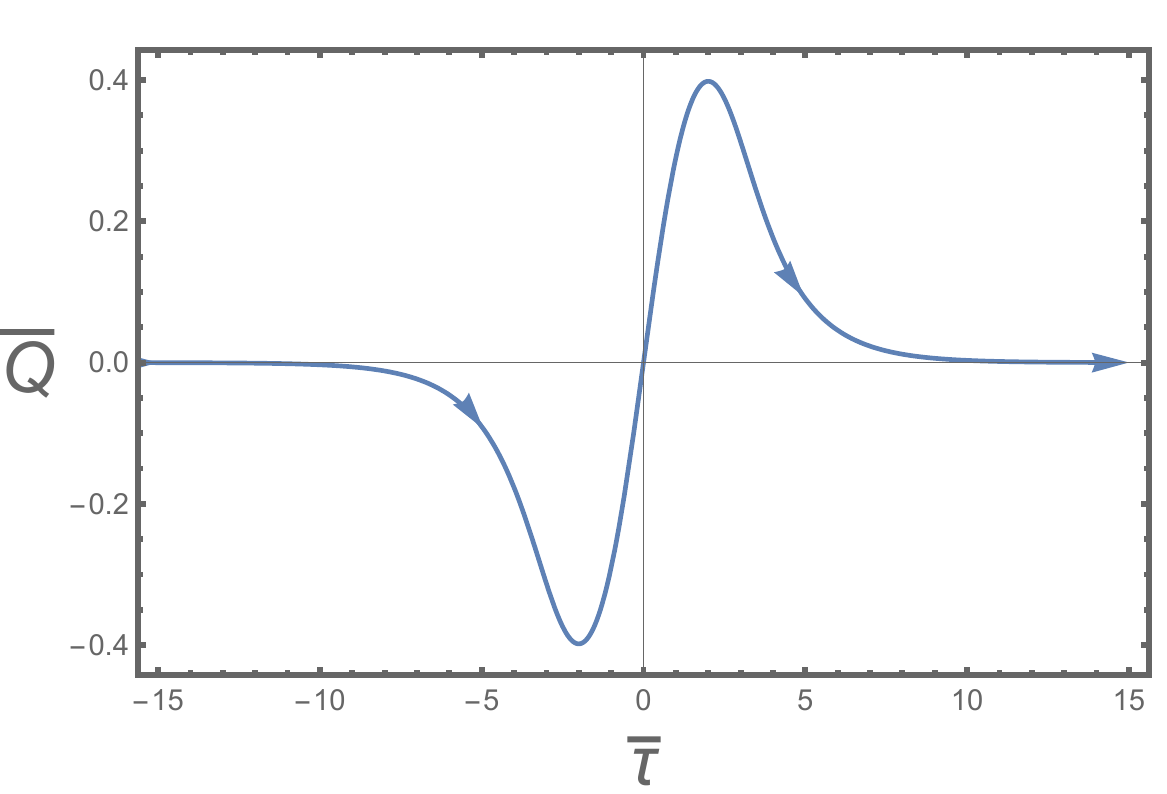}\vspace{1cm}
\caption{The left and right panels depict the temporal evolution of the variables $\Bar{x}$ and $\Bar{Q}$ for the variable $\Bar{\tau}$. }\label{2D_sys_plots_Ex3}
\end{figure}

Considering the spatially flat vacuum case, one can write the dynamical system as
\begin{subequations}
    \begin{eqnarray}
        \frac{d\Bar{x}}{d\Bar{\tau}}&=&\frac{1}{6}  \Big[2\left(-1+\Bar{x}\Bar{Q}+2\Bar{Q}^{2}-2\Bar{x}^{2}\Bar{Q}^{2}-\Bar{x}^{3}\Bar{Q}+\Bar{x}^{4}\right)+2 \Bar{x} \Bar{Q}(3 \Bar{Q}^2+\Bar{x}^2-1)\nonumber \\
        &&+2 \Bar{x} \Bar{Q} \sqrt{\left(\Bar{x}^2-1\right) \left(3 \Bar{Q}^2+\Bar{x}^2-1\right)+3\bar{Q}^4}\Big],\\   
        \frac{d\Bar{Q}}{d\Bar{\tau}}&=&\frac{1}{6} \Big[-4\Bar{Q}^{2}+2\Bar{x}\Bar{Q}+2\left(1-\Bar{x}^{2}\right) -2\Bar{x}^{2}\Bar{Q}^{2}+2\Bar{x}^{3}\Bar{Q}-4\Bar{x}\Bar{Q}^{3}+2 \Bar{Q}^{2}(3 \Bar{Q}^2+\Bar{x}^2-1)\nonumber \\
        &&+2 \bar{Q}^{2}\sqrt{\left(\bar{x}^2-1\right) \left(3 \bar{Q}^2+\bar{x}^2-1\right)+3\bar{Q}^4}\Big]\,,
    \end{eqnarray}    
\end{subequations}
whose fixed points are listed in Table \ref{tab:crit_model_III}.
\begin{table}[H]
    \centering
    \begin{tabular}{|c|c|c|c|c|}\hline
 \begin{tabular}{@{}c@{}} Fixed \\ points \end{tabular}  & \begin{tabular}{@{}c@{}} Coordinates\\ $(\Bar{x},\Bar{Q})$ \end{tabular} & Stability & Eigenvalues & Cosmology \\ \hline
  $A_{1\pm}$ & $(\pm1,0)$ & \begin{tabular}{@{}c@{}} $(-)$ attractor\\ $(+)$ repeller \end{tabular} & $\left\{\pm \frac{4}{3},\pm \frac{2}{3}\right\}$ & Indeterminate \\ \hline
    \end{tabular}
    \caption{Stability and cosmology of critical points for example III.}
    \label{tab:crit_model_III}
\end{table}
The corresponding phase portrait is shown in Figure \ref{fig:model_III}.
The time evolution of the dynamical variables are shown in Figure \ref{2D_sys_plots_Ex3}. 
We note that the cyclic nature of the evolution is not apparent from these plots, as these plots are generated using the numerical solutions of the dynamical system, which gives the fixed point structure strictly within one period $t\in\left[-\frac{\pi}{\zeta},\frac{\pi}{\zeta}\right]$. 

\section{Some discussion about the formulation} \label{sec:discussion}

In the section \ref{sec:applications} we have started our analysis with particular cosmographic ansatz e.g. the bouncing ansatz \eqref{eq:model-I}, \eqref{eq:model-II} or the cyclic ansatz \eqref{eq:model-III}. Nevertheless, we still get other possible cosmological evolutions as solutions of the system. For example, it is clear from Figs.\ref{fig:model_I},\ref{fig:model_II} that not all the phase trajectories correspond to a bounce, even though we originally started with a bouncing ansatz. Also, one can see that some of the fixed points we obtained for the second model correspond to a cosmological evolution of the form $a(t)=a_0+Ct$, which does not seem to be any asymptotic form of the original ansatz $a(t)=a_0\cosh{\lambda t}$. Naturally, one may ask why are we getting such solutions at all given that we have already started with an ansatz $a(t)$. We discuss the answer below, which also offers a deeper look at the underlying mechanism at play in our particular formulation.

In the usual dynamical system approach, one is given with a function $f(R)$. One then calculates the quantity $\Gamma\equiv\frac{F}{RF'}$ and expresses it in terms of the dynamical variables whenever possible, which helps in closing the autonomous system and proceed with the dynamical system analysis. In the ``cosmographic formulation'' of dynamical system that we have developed here for the purpose of a model-independent study, the quantity $\Gamma$ does not enter the dynamical system explicitly anymore. $\Gamma$ is given implicitly by Eqs.\eqref{eq:cond_a} and \eqref{eq:dec_non_vacuum} in terms of the dynamical variables $\bar{x},\bar{Q},\bar{K},\bar{\Omega}$ once a cosmology $j=j(q)$ is specified. Essentially, what our dynamical system formulation produces is the phase portrait of the underlying $f(R)$ whose explicit reconstruction is precisely what we chose to bypass via our method.

Take, for example, the ansatz \eqref{eq:model-II}. In the absence of any matter and for globally spatially flat cosmology, the Friedmann equation \eqref{eq:fried} can be written as a second order homogeneous differential equation of $f(R)$ with respect to $R$, whose general solution gives the most general form of the reconstructed $f(R)$. Following the classical reconstructed method (see, e.g. \cite[Sec(3.1)]{Carloni:2010ph}) we obtain the very complicated form \footnote{The form we obtained here apparently differs from the one obtained in \cite{Bouhmadi-Lopez:2012piq}, which follows a slightly different reconstruction procedure. In the absence of simple compact forms of the reconstructed $f(R)$, it is very difficult to compare analytically the functional forms $f(R)$ obtained via two different reconstruction routes. Also, the form presented in \cite[Eq.(2.15)]{Bouhmadi-Lopez:2012piq} is not the most general one, as the most general form will contain two integration constants.}
\begin{eqnarray}\label{eq:recon_model-II}
    f(R) = \left(-9 \lambda ^2+\sqrt{R-6 \lambda ^2} \sqrt{R-12 \lambda ^2}+R\right)^{\frac{\sqrt{3}}{2}} \sqrt{15 \sqrt{3} \lambda ^2+3 \sqrt{R-6 \lambda ^2} \sqrt{R-12 \lambda ^2}-2 \sqrt{3} R} \nonumber\\
    \left[\mathcal{C}_1 + \mathcal{C}_2 \int _1^{R}\frac{\sqrt{\chi-6 \lambda ^2} \left(\sqrt{\chi-12 \lambda ^2} \sqrt{\chi-6 \lambda ^2}+\chi-9 \lambda ^2\right)^{-\sqrt{3}}}{\sqrt{\chi-12 \lambda ^2} \left(3 \sqrt{\chi-12 \lambda ^2} \sqrt{\chi-6 \lambda ^2}-2 \sqrt{3} \chi+15 \sqrt{3} \lambda ^2\right)}d\chi\right]\,.
\end{eqnarray}

$\mathcal{C}_{1,2}$ are arbitrary constants. Although this $f(R)$ was reconstructed from the ansatz \eqref{eq:model-II}, this ansatz need not be the only possible solution admitted by the above $f(R)$. The above $f(R)$ can admit many other spatially flat vacuum cosmological scenarios as its solution, some of which are captured by our particular dynamical system approach. The phase portrait in Fig.\ref{fig:model_II} shows the spatially flat vacuum cosmological solution space of the above $f(R)$, which clearly shows it can also admit non-bouncing solutions. Indeed, the fixed points $A_{4\pm},\,A_{5\pm}$ which correspond to an evolution of the form $a(t)=a_0+Ct$ are cosmological phases along such non-bouncing trajectories. For the sake of completeness we have listed \emph{all} the fixed points. 

The same discussion holds for all the three models.

The complicated form of the reconstructed $f(R)$ in Eq.\eqref{eq:recon_model-II} actually serves to showcase the novelty of our approach. While considering nonsingular bouncing solutions in a theory, one is typically interested in questions like
\begin{itemize}
    \item Is a bouncing solution stable with respect to small homogeneous perturbations? If one starts with a particular bouncing trajectory and jumps to a nearby trajectory some time during the contracting part of the phase space, does one still end up with a bounce?
    \item Is a bouncing solution stable with respect to small inhomogeneous perturbations? Does it encounter ghost instability ($f'<0$) or tachyonic instability ($f''<0$) during its course of evolution?
\end{itemize}
It would be immensely hard to answer these questions in a straightforward manner with an absurdly complicated $f(R)$ form like the one in Eq.\eqref{eq:recon_model-II}. On the other hand, our approach answers these questions in an elegant manner while completely bypassing working with such complicated $f(R)$ forms. For spatially flat vacuum bounces, such questions can be answered particularly easily, by just looking at the corresponding 2D phase portraits. Also, to answer these questions, one can only focus on the bouncing trajectories and need not bother about other possible solutions of the general reconstructed $f(R)$ that may appear in the phase portrait.

\section{Conclusion} \label{sec:conclusion}
In this paper, a new model independent dynamical systems approach based on cosmographic parameters \cite{Chakraborty:2021jku} was developed further by compactifying the phase space for a general $f(R)$ theory of gravity and giving the corresponding physical viability conditions on the phase space.

In order to showcase the applicability of this approach, three examples displaying cyclic and/or bounce behaviour were studied. Fixed points as well as the associated stability for all systems were found. For the examples involving bouncing cosmology, it was found that the past and future asymptotic states of the bounce manifest themselves as invariant submanifolds in the phase space. The phase trajectories corresponding to bounces are confined within the region bounded by these submanifolds. An interesting observation made was that cyclic and bounce behaviour was only seen clearly when studying the parametric plots with positive spatial curvature initial conditions. This is consistent with a recent paper \cite{MacDevette:2022hts} in which it was found through the use of the explicit functional form of the $f(R)$ as well as the constraints on the compact phase space variables that the Hu-Sawicki model with $n = 1 = C_1$ displays no physical viable bounces in a spatially flat universe. The question remains whether this is a general statement, i.e., whether $f(R)$ models associated to a cyclic or bounce scale factor in FLRW cosmology can only display physically viable cyclic or bounce behaviour when the curvature is positive. 

All three examples showcased fixed points for which the limiting value of the quantity $\frac{\dot{H}}{H^2}$ is indeterminate; $\frac{\dot{H}}{H^2}\rightarrow\frac{0}{0}$ on approach to the fixed point. The cosmological evolution corresponding to these fixed points could not be determined. This showcases the disadvantage of compactifying the phase space.

It is known that the metric formalism of $f(R)$ gravity that we have worked with here is dynamically equivalent to Brans-Dicke theory with the vanishing Brans-Dicke parameter that belongs to the Horndeski class. There exists a so-called \emph{no-go theorem} about stable nonsingular cosmologies in Horndeski theories \cite{Libanov:2016kfc,Kobayashi:2016xpl}. Counterexamples also exist \cite{Banerjee:2018svi}. The no-go theorem is formulated considering spatially flat nonsingular FLRW cosmology in the absence of any additional hydrodynamic matter. In this regard, it is interesting to discuss how the results obtained regarding spatially flat vacuum nonsingular cosmologies in this paper compare to the assertion of the no-go theorem. Since the phase space of spatially flat vacuum cosmologies becomes 2-dimensional where we can clearly show the physically viable regions, our formalism provides an excellent way to investigate this issue. From figure \ref{fig:model_II} one can see that a spatially flat vacuum stable nonsingular bounce of the form \eqref{eq:model-II} is not at all possible in $f(R)$ gravity. On the contrary, figures \ref{fig:model_I} and \ref{fig:model_III} show that spatially flat vacuum stable bouncing cosmology of the form \eqref{eq:model-I} and cyclic cosmology of the form \eqref{eq:model-III} is in principle possible in $f(R)$ gravity, providing further counterexamples. However, the set of such counterexamples is of measure zero, i.e., one single trajectory, namely the one passing through the origin. Of course, when matter is present, the situation is different.

We again emphasize the limitation of the particular formulation presented in this paper. The success of the formulation crucially relies on being able to write a cosmographic relation $j=j(q)$. Although this is the case for the examples we considered here, it will not be the case in general. The cosmographic condition corresponding to a generic cosmological solution of an $f(R)$ theory will in general also contain snap and lerk parameter. From the approach taken to construct the compact dynamical system formulation of section \ref{sec:mod_ind_comp_dyn_sys}, it is not straightforward to extend it to consider a cosmology given by a condition of the form $l=l(q,j,s)$. One needs to come up with a different way to construct a compact dynamical system.

It would be interesting to extend this work by writing the system of equations in terms of redshift (or scale factor). This idea involves evolving the resulting equations and using the definition of the variables to plot the corresponding $f(R)$ as a function of the Ricci scalar. In other words, this new reconstruction method could be used to find the functional forms of $f(R)$ models associated with different cosmography as well as comparing them to $\Lambda$CDM and observational data. The present work provides a basis for reconstructing backgrounds that may be used in studies of large-scale structures, independent of gravitational theory.

Finally, it is worth pointing out that, while the cosmographic parameters have been used in our work to specify the expansion history representing a bounce or cyclic evolution, the same methods provide a powerful way to constrain $f(R)$ dark energy models using the cosmographic series during the epoch of structure formation. This was recently done in \cite{MacDevette:2024wpg}. It would be interesting to see if one could construct an algebraic cosmographic relation linking an early-time bounce with a late-time accelerated cosmic evolution. For the explicit reconstruction of $f(R)$ for one such background evolution, the reader can see \cite[section 5]{Bamba_2014}. Utilizing the constraints placed on the present-day values of the cosmographic parameters by various data sets, one can possibly infer about the existence and stability of such kind of cosmological solutions in $f(R)$ gravity.

\section*{Acknowledgement} SC acknowledges funding support from the NSRF via the Program Management Unit for Human Resources and Institutional Development, Research and Innovation (Thailand) [grant number B13F670063].	
	
\bibliographystyle{unsrt}
\bibliography{refs}

\end{document}